\documentclass[preprint,12pt]{elsarticle}
\usepackage[left=1 in,right=1 in, top=1 in, bottom=1 in]{geometry}
\usepackage{amsmath}

\usepackage{graphicx,color} 

\usepackage{amsmath,amsfonts,amssymb,theorem}
\usepackage{url,subfig,anysize,wrapfig}
\usepackage{multirow,tabularx,afterpage,fancyhdr}
\usepackage{cases}
\usepackage{epstopdf}

\usepackage[linewidth=1pt]{mdframed}
\usepackage{lipsum}

\usepackage{lineno}
\usepackage{placeins}

\usepackage{tikz}
\usetikzlibrary{calc}

\journal{arXiv}

\begin{document}

\begin{frontmatter}



\title{A Tube Dynamics Perspective Governing Stability Transitions: An Example Based on Snap-through Buckling}

\date{\today} 


\author{Jun Zhong\corref{cor1}\fnref{label1}}
\ead{junzhong@vt.edu}
\fntext[label1]{Engineering Mechanics Program, Virginia Tech, Blacksburg, VA 24061, USA}

\author{Lawrence N. Virgin\fnref{label2}}
\fntext[label2]{Department of Mechanical Engineering and Materials Science, Duke University, Durham, NC 27708, USA}

\author{Shane D. Ross\fnref{label1}}

\begin{abstract}

The equilibrium configuration of an engineering structure, able to withstand a certain loading condition, is usually associated with a local minimum of the underlying potential energy. However, in the nonlinear context, there may be other equilibria present, and this brings with it the possibility of a transition to an alternative (remote) minimum. That is, given a sufficient disturbance, the structure might buckle, perhaps suddenly, to another shape. This paper considers the dynamic mechanisms under which such transitions (typically via saddle points) occur. A two-mode Hamiltonian is developed for a shallow arch/buckled beam. The resulting form of the potential energy---two stable wells connected by rank-1 saddle points---shows an analogy with resonance transitions in celestial mechanics or molecular reconfigurations in chemistry, whereas here the transition corresponds to switching between two stable structural configurations. Then, from Hamilton's equations, the analytical equilibria are determined and linearization of the equations of motion about the saddle is obtained. After computing the eigenvalues and eigenvectors of the coefficient matrix associated with the linearization, a symplectic transformation is given which puts the Hamiltonian into normal form and simplifies the equations, allowing us to use the conceptual framework known as tube dynamics. 
The flow in the equilibrium region of phase space as well as the invariant manifold tubes in position space are discussed. Also, we account for the addition of damping in the tube dynamics framework, which leads to a richer set of behaviors in transition dynamics than previously explored.	
\end{abstract}

\begin{keyword}

potential energy \sep transients \sep tube dynamics \sep dynamic buckling \sep invariant manifolds \sep Hamiltonian


\end{keyword}

\end{frontmatter}

\newpage
\section{Introduction}

The nonlinear behavior of slender structures under loading is often dominated by a potential energy function that possesses a number of stationary points corresponding to various equilibrium configurations \cite{WiVi2016,collins2012isomerization}. Some are  stable (local minima, or `well'), some are  unstable (local maxima or `hill-top'), and some correspond to saddle points, i.e., a shape with opposite curvature in different directions, but still unstable, having both stable and unstable directions. Interestingly, although difficult to observe experimentally, it is these saddle points that can have a profound organizing effect on global trajectories in a dynamics context. Thus, under a nominally fixed set of loads or a given configuration we may have the situation in which a system is at rest in a position of stable equilibrium, but, given sufficiently large perturbation (input of energy) may transition to a remote stable equilibrium \cite{virgin2017geometric}, or even collapse completely \cite{das2009symmetry,das2009pull}. The path taken during this transition is associated with the least energetic route, and this will typically correspond to a passage close to a saddle point: it is easier to take a path around a mountain than going directly over its peak.

For a single mechanical degree of freedom the transition from one potential energy minimum to another is relatively unambiguous \cite{mann2009energy,Thompson1984}. We can think of a twin-well oscillator and how it has no choice but to pass over an intermediate hilltop in transitioning to an adjacent minimum. For high-order systems trajectories have many more possible paths. But a system with two mechanical degrees of freedom (configuration space), and thus a 4 dimensional phase space, offers an intermediate situation: compelling conceptual clarity (i.e., the potential energy can be thought of as a surface or landscape), but still retaining a wider range of potential behavior over and above the aforementioned single oscillator (i.e., multiple ways of traversing and perhaps escaping from one potential well to another).

For the two degree of freedom system, the analog of the hilltop is the saddle point of the potential energy surface. The linearized dynamics near such a point yield an oscillatory mode and an exponential mode, with both asymptotically stable and unstable directions. For energies slightly above the saddle point, there is a bottleneck to the energy surface \cite{NaRo2017,KoLoMaRo2000}. Transitions from one side of the bottleneck can be understood in terms of sets of trajectories which are bounded by topological cylinders. The transition dynamics, which has in some contexts been known as tube dynamics \cite{Conley1968, LlMaSi1985, OzDeMeMa1990, DeMeTo1991, DeLeon1992, Topper1997, KoLoMaRo2000, GaKoMaRo2005, GaKoMaRoYa2006, MaRo2006, KoLoMaRo2011}, has been developed for studying transitions between stable states (the potential wells) in a number of disparate contexts, and here it is applied to a structural mechanics situation in which snap-through buckling \cite{collins2012isomerization} is the key phenomenological transition. Conditions are determined whereby the initial energy imparted to the system is characterized in terms of subsequent escape from the initial potential well.

\section{The Paradigm: Snap-through of an Arch/Buckled Beam}

A classic example of a saddle-node bifurcation in structural mechanics is the symmetric snap-through buckling of a shallow arch, in an essentially co-dimension 1 bifurcation \cite{Thompson1984}. However, if the arch (or equivalently a buckled beam) is {\it not} shallow then the typical mechanism of instability is an asymmetric snap-through, requiring two modes (symmetric and asymmetric) for characterization, and the instability corresponds to a subcritical pitchfork bifurcation. In both of these cases the transition is sudden and associated with a fast dynamic jump, since there is no longer any locally available stable equilibrium. This behavior is generic regardless of boundary conditions and is also exhibited by the laterally-loaded buckled beam \cite{Murphy1996, Wiebe2013}. We shall focus on this latter example, for relative simplicity of introduction. The essential focus here is that the underlying potential energy of this system consists of two potential energy wells (the original unloaded equilibrium and the snapped-through equilibrium), an unstable hilltop (the intermediate, straight, unstable equilibrium) and two saddle-points. The symmetry of this system is broken by small geometric imperfections. The question is: {\it how does the system escape its local potential energy well} in a dynamical systems sense?

\begin{figure}[!b]
\begin{center}
\includegraphics[width=1.0\textwidth]{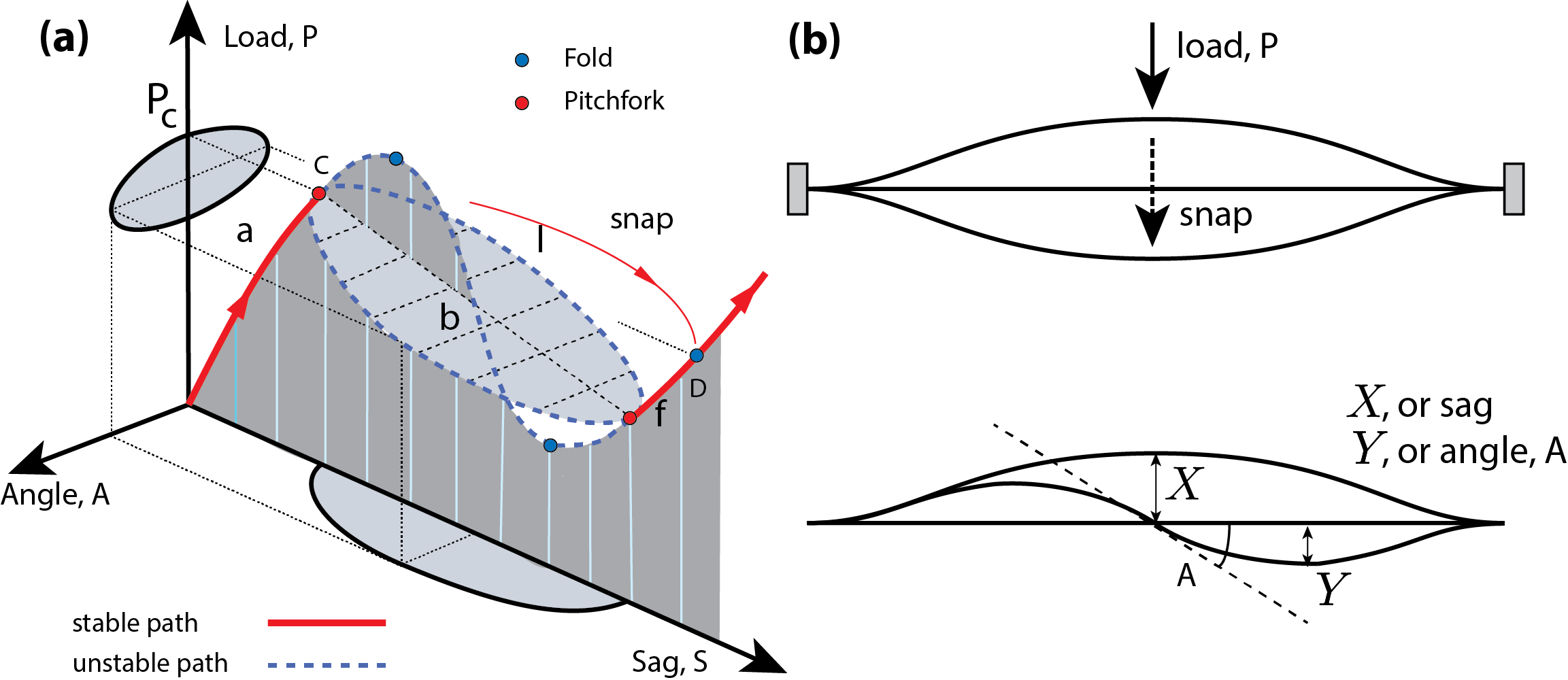}
\end{center}
\vspace{-4mm}
\caption{
\footnotesize
(a) A schematic load-deflection characteristic, (b) the two dominant degrees of freedom.}
\label{fig:arch}
\end{figure}

Suppose we have a moderately buckled beam. If a central point load is applied then the beam deflects initially in a purely symmetric mode, as shown by the red line in Fig.\ \ref{fig:arch}(a), following the $\alpha$ loading path.
Upon a quasi-static increase in the load $P$, point $C$ is reached (a subcritical pitchfork bifurcation) and the arch quickly snaps-through (a thoroughly dynamic event) with a significant asymmetric component in the deflection and the system settles into its inverted position $D$ \cite{virgin2017geometric}. This behavior is captured by considering a two-mode analysis: sag $S$ (symmetric) and angle $A$ (asymmetric), or alternatively we can use the harmonic coordinates $X$ and $Y$, respectively, corresponding to the modes in part (b). In an approximate analysis they might be the lowest two buckling modes or free vibration modes from a standard eigen-analysis. 
Suppose we load the beam to a value slightly below the snap value at $P_C$, and fix it at that value. In this case there will be the five equilibria mentioned earlier: three equilibria purely in sag (two stable and an unstable one between them), and two saddles, with significant angular components but geometrically opposed \cite{WiVi2016}. Small geometric imperfections (in $A$ and/or $S$) will break the symmetry and influence which path is more likely to be followed. In this fixed configuration we can then think of the system in dynamic terms, and consider the range of initial conditions (including velocity, perhaps caused by an impact force) that might push the system from a point on path $\alpha$ to a point on path $\phi$.

\paragraph{Governing equations} 
In this analysis a slender buckled beam with thickness $d$, width $b$ and length $L$ is considered. A Cartesian coordinate system $o \textendash xyz$ is established on the mid-plane of the beam in which $o$ is the origin, $x,y$ the directions along the length and width directions and $z$ the downward direction normal to the mid-plane. Based on Euler-Bernoulli beam theory \cite{zhong2016analysis,WiVi2016}, the displacement field $(u_1,u_3)$ of the beam along $(x,z)$ directions can be written as
\begin{equation}
	\begin{split}
		u_1(x,z,t)&= u(x,t)-z \frac{\partial w(x,t)}{\partial x}\\
		u_3(x,z,t)&= w(x,t)
		\label{disp_field}
	\end{split}
\end{equation}
where $u(x,t)$ and $w(x,t)$ are the axial and transverse displacements of an arbitrary point on the mid-plane of the beam. Considering the von K\'{a}m\'{a}n-type geometrical nonlinearity, the total axial strain can be obtained as
\begin{equation}
	\begin{split}
		\varepsilon^*_x= \frac{\partial u}{\partial x} - z \frac{\partial^2 w}{ \partial x^2}+ \frac{1}{2} \left(\frac{\partial w}{\partial x}\right)^2
		\label{total_strain}
	\end{split}
\end{equation}
For a moderately buckled-beam, we need to consider the initial strain $\varepsilon_0$ produced by initial deflection $w_0$ which is
\begin{equation}
	\begin{split}
		\varepsilon_0 = -z \frac{\partial^2 w_0}{ \partial x^2} + \frac{1}{2} \left(\frac{\partial w_0}{\partial x}\right)^2
		\label{init_strain}
	\end{split}
\end{equation}
Then the change in strain $\varepsilon_x$ can be expressed as 
\begin{equation}
	\begin{split}
		\varepsilon_x = \varepsilon^*_x - \varepsilon_0 = \frac{\partial u}{\partial x} - z \left(\frac{\partial^2 w}{\partial x^2} - \frac{\partial^2 w_0}{\partial x^2}\right) + \frac{1}{2} \left[ \left(\frac{\partial w}{\partial x}\right)^2 - \left(\frac{\partial w_0}{\partial x}\right)^2\right]
		\label{strain_change}
	\end{split}
\end{equation}

Here we just consider homogeneous isotropic materials with Young's modulus $E$, and allow for the possibility of thermal loading. The axial stress $\sigma_x$ can be obtained according to the one dimensional constitutive equation, as
\begin{equation}
	\begin{split}
		\sigma_x= E \varepsilon_x - E \alpha_x \Delta T
		\label{constitutive}
	\end{split}
\end{equation}
where $\alpha_x$ is  the thermal expansion coefficient and $\Delta T $ is the temperature increment from the reference temperature at which the beam is in a stress free state. Thermal loading is introduced as a convenient way of controlling the initial equilibrium shapes (and hence the potential energy landscape) via axial loading.

The strain energy $\mathcal{V}(x,z,t)$ is 
\begin{equation}
	\begin{split}
		\mathcal{V}(x,z,t) &= \frac{b}{2} \int_0^L \int_{- \frac{d}{2}}^{ \frac{d}{2}} \sigma_x \varepsilon_x \mathrm{d}z \mathrm{d}x
		\label{strain_energy}
	\end{split}
\end{equation}
Ignoring the axial inertia term, the kinetic energy $\mathcal{T}(x,z,t)$ of the buckled beam is
\begin{equation}
	\begin{split}
		\mathcal{T}(x,z,t)= \frac{b}{2} \int_0^L \int_{- \frac{d}{2}}^{ \frac{d}{2}} \rho \dot w^2  \mathrm{d}z \mathrm{d}x
		\label{kinetic_energy}
	\end{split}
\end{equation}
where $\rho$ is the mass density. In addition, the dot over the quantity is the derivative with respective to time.

The governing equations can be obtained by Hamilton's principle which requires that
\begin{equation}
	\begin{split}
		\delta \int_{t_0}^t \left[ \mathcal{T}(x,z,t) - \mathcal{V}(x,z,t) \right] \mathrm{d} t
		\label{Hamilton_prin} + \int_{t_0}^{t} \delta W_{nc} \mathrm{d}t =0
	\end{split}
\end{equation}
where $\delta$ denotes the variational operator, $t_0$ and $t$ the initial and current time. $\delta W_{nc}$ is the variation of the virtual work done by non-conservative force (damping) which is expressed as
\begin{equation}
	\begin{split}
		\delta W_{nc} = - c_d \dot w \delta w
	\end{split}
\end{equation} 
where $c_d$ is the coefficient of (linear viscous) damping. In subsequent analysis, and related to typical practical situations, the damping will be small.

After some manipulation, the governing equations in terms of axial force $N_x$ and bending moment $M_x$ can be obtained as \cite{zhong2016analysis}
\begin{equation}
	\begin{split}
		&\frac{\partial N_x}{\partial x}=0\\
		&\frac{\partial^2 M_x}{\partial x^2}+ N_x \frac{\partial^2 w}{\partial x^2} = \rho A \ddot w + c_d \dot w
		\label{govern_eq_force}
	\end{split}
\end{equation}
where $N_x$ and $M_x$ are defined as
\begin{equation}
	\begin{split}
		\left(N_x, M_x \right)=b \int_{- \frac{d}{2}}^{ \frac{d}{2}}
	\end{split} \sigma_x \left(1,z \right) \mathrm{d} z
	\label{force_definition}
\end{equation}
By using \eqref{disp_field}, \eqref{strain_change} and \eqref{constitutive}, the force $N_x$ and moment $M_x$ in \eqref{force_definition} can be rewritten as
\begin{equation}
	\begin{split}
		N_x&=E A \left[\frac{\partial u}{\partial x} + \frac{1}{2} \left( \left(\frac{\partial w}{\partial x}\right)^2 - \left(\frac{\partial w_0}{\partial x}\right)^2\right) \right] -N_T\\
		M_x&=- E I  \left(\frac{\partial^2 w}{\partial x^2} - \frac{\partial^2 w_0}{\partial x^2}\right)
		\label{force_moment}
	\end{split}
\end{equation}
where $A$ and $I$ denote the cross-sectional area and moment of inertia; $N_T=EA \alpha_x \Delta T$, the axial thermal loads. Thus, $EA$ and $EI$ are the axial stiffness and bending stiffness, respectively.

Here we just consider a clamped-clamped beam with in-plane immovable ends. The boundary conditions are 
\begin{equation}
	\begin{split}
		x=0,L: u=w=\frac{\partial w}{\partial x}=0
		\label{boundary_condi}
	\end{split}
\end{equation}
Note that from the first equation in \eqref{govern_eq_force}, it is clear that the axial force $N_x$ is constant along the axial direction. In this case, integrating the axial force along the $x$ axis and using the boundary conditions $u(0,t)=u(L,t)=0$, one can obtain
\begin{equation}
	\begin{split}
		N_x= \frac{E A}{2L} \int_0^L \left[ \left(\frac{\partial w}{\partial x} \right)^2 - \left(\frac{\partial w_0}{\partial x} \right)^2 \right] \mathrm{d} x -N_T
		\label{axial_force}
	\end{split}
\end{equation}

Using $M_x$ in \eqref{force_moment} and $N_x$ in \eqref{axial_force}, the second equation in \eqref{govern_eq_force} can be described in terms of the transverse displacement $w$ as \cite{WiVi2016}
\begin{equation}
       \begin{split}
              \rho A \ddot{w} + c_d \dot w + EI \left(\frac{\partial^4 w}{\partial x^4} - \frac{\partial^4 w_0}{\partial x^4}\right) + \left[N_T - \frac{EA}{2L}  \int_0^L \left( \left(\frac{\partial w}{\partial x} \right)^2 - \left(\frac{\partial w_0}{\partial x} \right)^2 \right) \mathrm{d} x \right] \frac{\partial^2 w}{\partial x^2} =0
              \label{eq:PDE}
       \end{split}
\end{equation}
where $w$ and $w_0$ are the current deflection and initial geometrical imperfection, respectively; $\rho$ is the mass density; $c_d$ is the damping coefficient; $A$ and $I$ are the area and the moment of inertia of the cross-section, respectively; $E$ is the Young's modulus. Given the immovable ends it is natural to consider the effective externally applied axial force to be replaced by a thermal loading term: this is the primary destabilizing nonlinearity in the system.

\begin{figure}[!b]
\begin{center}
\includegraphics[width=1.0\textwidth]{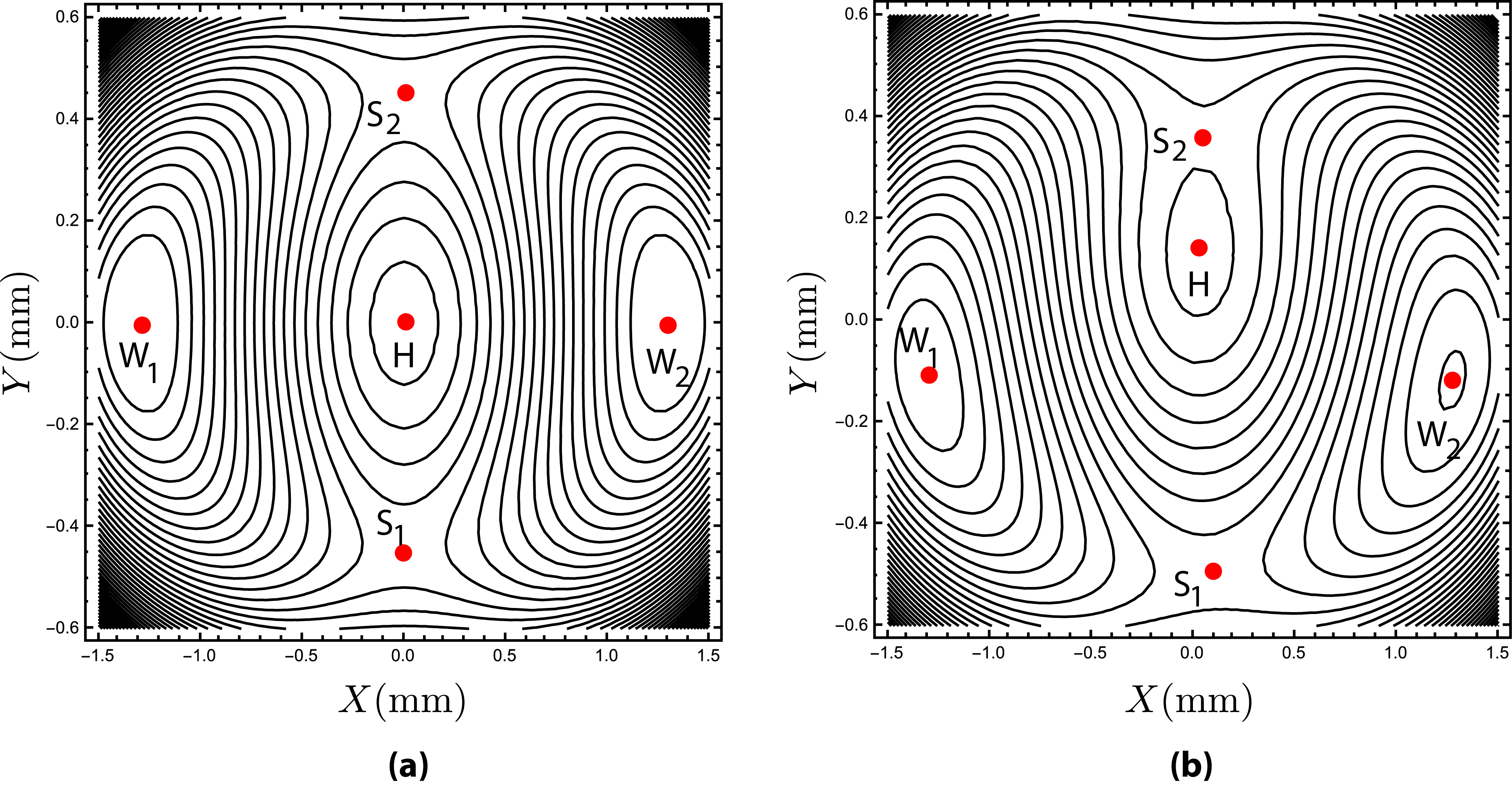}
\end{center}
\vspace{-4mm}
\caption{
\footnotesize
Contours of potential energy: (a) the symmetric system, $\gamma_1 = \gamma_2 = 0$, (b) with small initial imperfections in both modes, i.e., $\gamma_1$ and $\gamma_2 $ are nonzero.}
\label{fig:shape}
\end{figure}

As mentioned earlier, clamped-clamped boundary conditions are considered. Thus we make use of the mode shapes
\begin{equation}
      \begin{split}
            &\phi_n = \alpha_n \left[\sinh  \frac{\beta_n x}{L} - \sin  \frac{\beta_n x}{L} + \delta_n \left(\cosh  \frac{\beta_n x}{L} - \cos  \frac{\beta_n x}{L} \right) \right],\\
            &\delta_n = \frac{\sinh \beta_n - \sin \beta_n}{\cos \beta_n - \cosh \beta_n},\\
            &\cos \beta_n \cosh \beta_n =1,\\
            & \alpha_1 = -0.6186, \ \ \  \alpha_2 = -0.6631
            \label{mode:Virgin}
      \end{split}
\end{equation}
and describe the deflected shape in terms of a two-degree-of-freedom approximation
\begin{equation}
       \begin{split}
              w(x,t)&= X(t) \phi_1(x) +Y(t) \phi_2(x),\\
              w_0(x)&= \gamma_1 \phi_1(x) + \gamma_2 \phi_2(x)
       \end{split}
\end{equation}
where the initial imperfections are given by $w_0$.
Substituting the assumed solution into the equation of motion ~\ref{eq:PDE} yields
\begin{equation}
       \begin{split}
              &\rho A \int_0^L \phi_i \ddot w \mathrm{d}x + c_d \int_0^L \phi_i \dot w \mathrm{d}x+ EI \int_0^L \frac{\partial^2 \phi_i}{\partial x^2} \left(\frac{\partial^2 w}{\partial x^2} - \frac{\partial^2 w_0}{\partial x^2} \right) \mathrm{d} x\\
              & - \left[N_T - \frac{EA}{2L} \int_0^L \left(\left(\frac{\partial w}{\partial x} \right)^2 - \left(\frac{\partial w_0}{\partial x} \right)^2 \right) \mathrm{d}x \right] \int_0^L \frac{\partial \phi_i}{\partial x} \frac{\partial w}{\partial x} \mathrm{d}x =0
              \label{virtual}
       \end{split}
\end{equation}
Using the specific forms of $\phi_i$ in \eqref{mode:Virgin} and noticing each mode shape is orthogonal, the nonlinear equations can be obtained
\begin{equation}
       \begin{split}
              & M_1 \ddot X + C_1 \dot X + K_1 \left(X - \gamma_1 \right) - N_T G_1 X - \frac{EA}{2L}G_1^2 \left(\gamma_1^2 X -X^3 \right) - \frac{EA}{2L} G_1 G_2 \left(\gamma_2^2 X -X Y^2 \right)=0\\
              & M_2 \ddot Y + C_2 \dot Y + K_2 \left(Y - \gamma_2 \right) - N_T G_2 Y - \frac{EA}{2L}G_2^2 \left(\gamma_2^2 Y -Y^3 \right) - \frac{EA}{2L} G_1 G_2 \left(\gamma_1^2 Y -X^2 Y \right)=0
              \label{odes}
       \end{split}
\end{equation}
where
\begin{equation}
       \begin{split}
              \left(M_i, C_i \right) = \left(\rho A, c_d \right) \int_0^L \phi_i^2 \mathrm{d}x, \ \ K_i = EI \int_0^L \left(\frac{\partial^2 \phi_i}{\partial x^2} \right)^2 \mathrm{d}x, \ \ G_i = \int_0^L \left(\frac{\partial \phi_i}{\partial x} \right)^2 \mathrm{d}x
              \label{Galerkin-coefficient}
       \end{split}
\end{equation}
The kinetic energy and potential energy, respectively, can be represented as
\begin{equation}
       \begin{split}
              \mathcal{T}(\dot X, \dot Y)=& \frac{1}{2} M_1 \dot X^2 + \frac{1}{2} M_2 \dot Y^2,\\
              \mathcal{V}(X, Y)=& - K_1 \gamma_1 X - K_2 \gamma_2 Y + \frac{1}{2} K_1 X^2 + \frac{1}{2} K_2 Y^2 - \frac{1}{2} N_T\left( G_1 X^2 +  G_2 Y^2 \right) \\
              & - \frac{EA}{2L}G_1^2 \left(\frac{1}{2}\gamma_1^2 X^2 - \frac{1}{4}X^4 \right) - \frac{EA}{2L}G_2^2 \left(\frac{1}{2} \gamma_2^2 Y^2 - \frac{1}{4}Y^4 \right) \\
              & - \frac{EA}{2L} \frac{G_1 G_2}{2} \left(\gamma_2^2 X^2 + \gamma_1^2 Y^2 -X^2 Y^2 \right).\\
       \end{split}
\end{equation}
For physically reasonable coefficients we have a number of equilibrium possibilities. For small values of $N_T$ we have an essentially linear system, dominated by the trivial (straight) equilibrium configuration, and thus an isolated center (minimum of the potential energy). This relates back to the situation in Figure~\ref{fig:arch} for a small value of $P$. But for larger values of $P$, for example a little below $P_c$, the system typically possesses a number of equilibria, some of which are stable and some of which are not. Some typical forms are shown in Figure~\ref{fig:shape}(a) in which the five dots are the equilibrium points where W$_1$ and W$_2$ are within the two stable wells; S$_1$ and S$_2$  two unstable saddle points; H the unstable hilltop. Thus, we might have the system sitting (in equilibrium) at point W$_1$. If it is then subject to a disturbance {\it with the right size and direction} (in the dynamical context), we might expect the system to transition to the remote equilibrium at W$_2$. This might occur when the system is subject to a large impact force, for example \cite{Wiebe2013}. It is anticipated (and will later be shown) that the typically easiest transition will be associated with (an asymmetric) passage close to S$_1$ or S$_2$, and generally avoiding H. In Figure~\ref{fig:shape}(b) is shown the same system but now with a small geometric imperfection in both modes (i.e., $\gamma_1 \ne 0$ and $\gamma_2 \ne 0$). In this case the symmetry of the system is broken, and given the relative energy associated with the saddle points it is anticipated (and will also be shown later) that optimal escape will tend to occur via S$_1$.

Note that eqs.\ \eqref{odes} can also be obtained from Lagrange's equations,
\begin{equation}
\frac{d}{dt}\left( \frac{\partial \mathcal{L}}{ \partial \dot q_i}\right) -\frac{\partial \mathcal{L}}{ \partial  q_i} = - C_i \dot q_i, \quad i =1, 2
\end{equation}
when $q_1 = X$ and $q_2=Y$, and the Lagrangian is
\begin{equation}
\mathcal{L}(X,Y,\dot X,\dot Y) = \mathcal{T}(\dot X, \dot Y) - \mathcal{V}(X, Y)
\end{equation}

To transform this to a Hamiltonian system, one defines the generalized momenta,
\begin{equation}
\begin{split}
p_i = \frac{\partial \mathcal{L}}{ \partial \dot q_i} = M_i \dot q_i
\end{split}
\end{equation}
so $p_X = M_1 \dot X$ and $p_Y = M_2 \dot Y$,
in which case, the kinetic energy is
\begin{equation}
\mathcal{T}(X,Y,p_X,p_Y)  = \frac{1}{2 M_1} p_X^2 +  \frac{1}{2 M_2} p_Y^2 
\end{equation}
and the Hamiltonian is 
\begin{equation}
\mathcal{H}(X,Y,p_X,p_Y) = \mathcal{T} + \mathcal{V}
\end{equation}
and Hamilton's equations (with damping) \cite{Greenwood2003} are
\begin{equation}
\begin{split}
\dot X &= \frac{\partial \mathcal{H}}{\partial p_X}=\frac{p_X}{M_1}  \\
\dot Y &= \frac{\partial \mathcal{H}}{\partial p_Y}=\frac{p_y}{M_2} \\
\dot p_X &= - \frac{\partial \mathcal{H}}{\partial X} - C_H p_X=- \frac{\partial \mathcal{V}}{\partial X} - C_H p_X \\
\dot p_Y &= - \frac{\partial \mathcal{H}}{\partial Y} - C_H p_Y=- \frac{\partial \mathcal{V}}{\partial Y} - C_H p_Y \\
\label{eq:eomHam}
\end{split}
\end{equation}
where
\begin{equation}
\begin{split}
 \frac{\partial \mathcal{V}}{\partial X}=& K_1 \left(X - \gamma_1 \right) - N_T G_1 X - \frac{EA}{2L}G_1^2 \left(\gamma_1^2 X -X^3 \right) - \frac{EA}{2L} G_1 G_2 \left(\gamma_2^2 X -X Y^2 \right),\\
 \frac{\partial \mathcal{V}}{\partial Y}=& K_2 \left(Y - \gamma_2 \right) - N_T G_2 Y - \frac{EA}{2L}G_2^2 \left(\gamma_2^2 Y -Y^3 \right) - \frac{EA}{2L} G_1 G_2 \left(\gamma_1^2 Y -X^2 Y \right)
\end{split}
\end{equation}
and $C_H=C_1/M_1=C_2/ M_2$ is the damping coefficient in the Hamiltonian system which can be easily found by comparing \eqref{odes} and \eqref{eq:eomHam}, and using the relations of $M_i$ and $C_i$ in  \eqref{Galerkin-coefficient}.
	
We assume the lower saddle point S$_1$ has the smaller potential energy compared to S$_2$, thus the energy of S$_1$ is the critical energy for snap-though, and we initially focus attention on the dynamic behavior around the region of S$_1$. The linearized equations of  \eqref{eq:eomHam} about S$_1$ with position $(X_e,Y_e)$ can be written as
\begin{equation}
	\begin{split}
		\dot x&= \frac{p_x}{M_1}\\
		\dot y&= \frac{p_y}{M_2}\\
		\dot p_x&= A_{31} x + A_{32} y - C_H p_x\\
		\dot p_y&= A_{32} x + A_{42} y - C_H p_y
		\label{linearization}
	\end{split}
\end{equation}
where $(x,y,p_x,p_y)= (X,Y,p_X,p_Y) - (X_e,Y_e,0,0)$ and
\begin{equation}
      \begin{split}
       & A_{31}= -K_1 + N_T G_1 + \frac{E A G_1^2 \left(\gamma_1^2 - 3 X_e^2 \right)}{2L} + \frac{E A G_1 G_2 \left(\gamma_2^2 -Y_e^2 \right)}{2L},\\
       & A_{32}= - \frac{E A G_1 G_2 X_e Y_e}{L},\\
       & A_{42}= -K_2 + N_T G_2 + \frac{E A G_2^2 \left(\gamma_2^2 - 3 Y_e^2 \right)}{2L} + \frac{E A G_1 G_2 \left(\gamma_1^2 -X_e^2 \right)}{2L}
       \label{lin paras}
        \end{split}
\end{equation}

If we replace the position of S$_1$ by the position of W$_1$, we can still use the linearized equations in \eqref{linearization} to obtain the natural frequencies of the shallow arch near W$_1$ as
\begin{equation}
      \omega_{1,2}^{(d)} = w_{1,2}^{(c)} \sqrt{1-\xi_{1,2}^2}
\end{equation} 
where $\omega_{1,2}^{(c)}$ are the first two natural frequencies for the conservative system and $\xi_{1,2}$ are the viscous damping factors with the forms
\begin{equation}
\omega_{1,2}^{(c)}=\frac{(b_{\omega} \mp \sqrt{b_{\omega}^2 - 4 c_{\omega}} )}{2}, \hspace{0.5in} \xi_{1,2} = \frac{C_H}{2 \omega_{1,2}^{(c)}}
\end{equation}
and
\begin{equation*}
     b_{\omega}=- \frac{A_{31}}{M_1} - \frac{A_{42}}{M_2}, \hspace{0.5in} c_{\omega}=\frac{A_{31} A_{42} - A_{32}^2}{M_1 M_2}
\end{equation*}

\paragraph{Non-dimensional equations of motion}
In order to reduce the parameters, some non-dimensional quantities are introduced,
\begin{equation}
       \begin{split}
              &\left( L_x, L_y \right)= L  \left(1,\sqrt{ \frac{M_1}{M_2}} \right), \omega_0= \frac{ \sqrt{- A_{32} }}{ \left( M_1 M_2\right)^ \frac{1}{4}}, \tau= \omega_0 t,  \left(\bar q_1 , \bar q_2 \right)= \left( \frac{x}{L_x}, \frac{y}{L_y} \right),\\
              &\left(\bar p_1 , \bar p_2 \right)= \frac{1}{\omega_0} \left(  \frac{p_x}{ L_x M_1},\frac{p_y}{ L_y M_2}\right), \left( c_x  , c_y \right)= \frac{1}{ \omega_{0}^2} \left( \frac{A_{31}}{M_1}, \frac{A_{42}}{M_2} \right), c_1= \frac{C_H}{\omega_0}
              \label{dimless quan}
       \end{split}
\end{equation}

Using the non-dimensional parameters in \eqref{dimless quan}, the non-dimensional linearized equations are written as
\begin{equation}
\begin{split}
\dot {\bar q}_1 &= \bar p_1,\\
\dot {\bar q}_2 &= \bar p_2,\\
\dot {\bar p}_1 &= c_x \bar q_1 - \bar q_2 - c_1 \bar p_1,\\
\dot {\bar p}_2 &=  -  \bar q_1 + c_y \bar q_2 - c_1 \bar p_2
\label{nond eq}
\end{split}
\end{equation}
Written in matrix form, with column vector $\bar z=(\bar q_1 , \bar q_2 , \bar p_1 , \bar p_2)$, we have
\[
\dot {\bar z} = A \bar z + D \bar z
\]
where
\begin{equation}
A = \begin{pmatrix}
0     	 & 0 & 1 & 0 \\
0	  & 0 & 0 & 1 \\
c_x	  & -1 & 0 & 0 \\
-1	  & c_y & 0 & 0 
\end{pmatrix},
\hspace{0.5in}
D = \begin{pmatrix}
0     	 & 0 & 0 & 0 \\
0	  & 0 & 0 & 0 \\
0	  & 0 & -c_1 & 0 \\
0	  & 0 & 0 & -c_1 
\end{pmatrix}
\label{A_and_D_matrix}
\end{equation}
are the Hamiltonian part and damping part of the linear equations, respectively. 

\section{Linearized Conservative Hamiltonian System} \label{linearization of conserve}

\subsection{Solutions near the equilibria}

\paragraph{Eigenvalues and eigenvectors} 
In this section, we will discuss the linear dynamical behaviors of a buckled beam in the Hamiltonian system without taking account of any energy dissipation which makes $c_1=0$ (i.e., $C_H = 0$). Thus, the equations of motion are given as
\begin{equation}
\dot {\bar z} = A \bar z
\label{conservative eqns}
\end{equation}
The system \eqref{conservative eqns} can be viewed as resulting from a quadratic Hamiltonian,
\begin{equation}
\mathcal{H}_2= \tfrac{1}{2}\bar p_1 ^2 + \tfrac{1}{2}\bar p_2 ^2 
- \tfrac{1}{2}c_x \bar q_1 ^2 - \tfrac{1}{2}c_y \bar q_2 ^2 + \bar q_1 \bar q_2
\label{H_2_bar}
\end{equation}
which can be written in matrix form
\[
\mathcal{H}_2 = \frac{1}{2} \bar z ^T B \bar z
\]
where
\begin{equation*}
\begin{split}
B = J^{T}A =\begin{pmatrix}
-c_x      & 1 & 0 & 0 \\
1	  & -c_y & 0 & 0 \\
0	  & 0 & 1 & 0 \\
0	  & 0 & 0 & 1 
\end{pmatrix}
\end{split}
\end{equation*}
and ${J}$ is the $4 \times 4$ canonical symplectic matrix
\begin{eqnarray*}
\begin{split}
{J} = \begin{pmatrix}
{\bf 0} & {I}_2\\
-{I}_2 & {\bf 0}\\
\end{pmatrix}
\end{split}
\end{eqnarray*}
where ${I}_2$ is the $2 \times 2$ identity matrix.

The characteristic polynomial of \eqref{conservative eqns} is 
\begin{align*}
       p( \beta ) = \beta^4 - ( c_x + c_y ) \beta^2 + c_x c_y - 1
\end{align*}
Let $\alpha= \beta^2$, then the roots of $ p(\alpha)=0 $ are as follows
\begin{equation}
        \begin{split}
              \alpha_1= \frac{c_x + c_y  + \sqrt{ \left( c_x -c_y  \right)^2 + 4}}{2},\\ \alpha_2= \frac{c_x + c_y  - \sqrt{ \left( c_x - c_y  \right)^2 + 4}}{2}
              \label{charoots}
        \end{split}
\end{equation}

Generally, in \eqref{conservative eqns} $c_x > 0$ and $c_y <0$. In this case, $\alpha_1 > 0$ and $\alpha_2 < 0$. It follows that this equilibrium point is of the type saddle $\times$ center. Here we define $\lambda=\sqrt{\alpha_1}$ and $\omega_p=\sqrt{-\alpha_2}$. Thus, the eigenvectors are given by
\begin{equation}
         \begin{split}
                \left(1, c_x - \beta^2 , \beta , c_x \beta -\beta^3 \right),
                \label{gener egvec}
         \end{split}
\end{equation}
where $\beta$ denotes one of the eigenvalues.

After substituting $\beta = i \omega_p $ into \eqref{gener egvec} and separating real and imaginary parts as $u_{\omega_p} + i v_{\omega_p}$, we obtain two corresponding eigenvectors
\begin{equation}
         \begin{split}
                u_ {\omega_p}&= \left( 1, c_x+ \omega_p^2 , 0 , 0 \right),\\
                v_ {\omega_p}&= \left( 0, 0 , \omega_p ,c_x \omega_p+ \omega_p^3  \right),
                \label{eivect I}
         \end{split}
\end{equation}

Moreover, the other two eigenvectors associated with the pair of real eigenvalues $\pm \lambda $ can be taken as 
\begin{equation}
         \begin{split}
                 u_{+ \lambda}&= \left( 1, c_x - \lambda^2 , \lambda , c_x \lambda - \lambda^3 \right),\\
                 u_{- \lambda}&=- \left( 1, c_x - \lambda^2 , - \lambda , \lambda^3 - c_x \lambda \right)
                 \label{eivect R}
         \end{split}
\end{equation}

\paragraph{Symplectic change of variables}
We consider the linear symplectic change of variables from $(\bar q_1, \bar q_2, \bar p_1, \bar p_2)$ to $(q_1 , q_2 , p_1 , p_2)$,
\begin{equation}
 \left( \begin{array}{c c c c} \bar q_1 \\ \bar q_2 \\ \bar p_1 \\ \bar p_2
\end{array} 
 \right) = 
C
\left( 
\begin{array}{c c c c} q_1 \\ q_2 \\ p_1 \\ p_2
\end{array}  
\right) 
\end{equation} \label{CT}
where the columns of the matrix $C$ are given by the eigenvectors,
\begin{equation}
        \begin{split}
               C=\left(u_{ + \lambda} , u_{ \omega_p} , u_{ - \lambda} , v_{\omega_p} \right)
               \label{change matrix}
        \end{split}
\end{equation}
and where the vectors are written as column vectors.

Then we find
\begin{equation}
        \begin{split}
               C^T J C=\begin{pmatrix}
                0 & \bar D\\
                -\bar D & 0 
               \end{pmatrix}, \hspace{0.5in}
               \bar D= \begin{pmatrix}
               d_\lambda & 0\\
               0 & d_{\omega_p}
               \end{pmatrix}
        \end{split}
\end{equation}
where
\begin{equation}
        \begin{split}
               d_\lambda &= \lambda [4 - 2 (c_x -c_y) ( \lambda^2 -c_x)]\\
               d_{\omega_p}&= \frac{\omega_p}{2} [4 +2 (c_x - c_y) ( \omega_p^2 + c_x)]
        \end{split}
\end{equation}

In order to obtain a symplectic form which satisfies $C^T J C =J$, we need to rescale the columns of $C$. The scaling is given by factors $s_1 = \sqrt{ d_\lambda}$ and $s_2 = \sqrt{d_{\omega_p}}$. In this case, the final form of the symplectic matrix $C$ is given by
\begin{equation}
       \begin{split}
              C= \begin{pmatrix}
              \frac{1}{s_1} & \frac{1}{s_2} & - \frac{1}{s_1} & 0\\
              \frac{c_x - \lambda^2}{s_1} & \frac{ \omega_p^2 + c_x}{s_2} & \frac{\lambda^2 - c_x}{s_1}  & 0\\
              \frac{\lambda}{s_1} & 0 & \frac{ \lambda}{s_1} & \frac{\omega_p}{s_2} \\
              \frac{c_x  \lambda - \lambda^3}{s_1} & 0 & \frac{c_x \lambda - \lambda^3}{s_1} &  \frac{c_x \omega_p + \omega_p^3 }{s_2} 
              \end{pmatrix}
              \label{sym matrix}
       \end{split}
\end{equation}

The Hamiltonian \eqref{H_2_bar} can be rewritten in the simplified, normal form,
\begin{equation}\label{Enlin}
        \mathcal{H}_2= \lambda q_1 p_1 + \tfrac{1}{2} \omega_p(q_2 ^2 +p_2^2)
\end{equation}
with corresponding linearized equations, 
\begin{equation}
         \begin{split}
           \dot q_1&= ~~\lambda q_1, \\
           \dot p_1&= - \lambda p_1,\\
           \dot q_2&=  ~~\omega_p p_2, \\
            \dot p_2&= - \omega_p q_2
           \label{new equa}
         \end{split}         
\end{equation}
Written in matrix form, with column vector $z=(q_1 , q_2 , p_1 , p_2)$, we have
\[
\dot z = \Lambda z
\]
where
\begin{equation}
       \begin{split}
              \Lambda = C^{-1}AC = \left( \begin{array}{rrrr}
              \lambda & 0 & 0 & 0 \\
               0	  & 0 & 0 & \omega_p \\
                 0	  & 0 & -\lambda & 0 \\
                 0	  & -\omega_p & 0 & 0 
              \end{array} \right)
              \label{Lambda_matrix}
       \end{split}
\end{equation}

The solution of \eqref{new equa} can be written as
\begin{equation}
        \begin{split}
              & q_1= q_1^0 e^{ \lambda t}, \ \ \ p_1= p_1^0 e^{ - \lambda t}\\
              & q_2 + i p_2 = \left(q_2^0 + i  p_2^0 \right) e^{-i \omega_p t}
        \end{split}
\end{equation}
Note that the three functions
\[
f_1 = q_1 p_1, \quad
f_2 =  q_2^2 + p_2^2, \quad
f_3 = \mathcal{H}_2 
\]
are constants of motion under the Hamiltonian system \eqref{new equa}.

\subsection{Boundary of transit and non-transit orbits}
\label{sec:separatrix}

\paragraph{The Linearized Phase Space} 
For positive $h$ and $c$,
the equilibrium or bottleneck region $\mathcal{R}$ (sometimes just called the neck region), which is determined by
\[
\mathcal{H}_2=h, \quad  \mbox{and} \quad |p_1-q_1|\leq c,
\]
is homeomorphic to the product of a 2-sphere and an interval $I$, 
$S^2\times I$;
namely, for each fixed value of $p_1 -q_1 $ in the interval $I=[-c,c]$,
we see that  the equation $\mathcal{H}_2=h$ determines a 2-sphere
\begin{equation}\label{2-sphere}
\tfrac{\lambda }{4}(q_1 +p_1 )^2
+ \tfrac{1}{2}\omega_p (q_2^2+p_2^2)
=h+\tfrac{\lambda }{4}(p_1 -q_1 )^2.
\end{equation}
Suppose $a \in I$, then \eqref{2-sphere} can be re-written as
 \begin{equation}\label{2-sphere2}
x_1^2 + q_2^2+p_2^2
= r^2,
\end{equation}
where $x_1 = \sqrt{\tfrac{1 }{2}\tfrac{\lambda}{\omega_p}}(q_1 +p_1 )$ and 
$r^2=\tfrac{2}{\omega_p}(h+\tfrac{\lambda }{4}a^2)$, which defines a 2-sphere of radius $r$ in the three variables $x_1$, $q_2$, and $p_2$.

\begin{figure}[ht]
\begin{center}
\includegraphics[width=\textwidth]{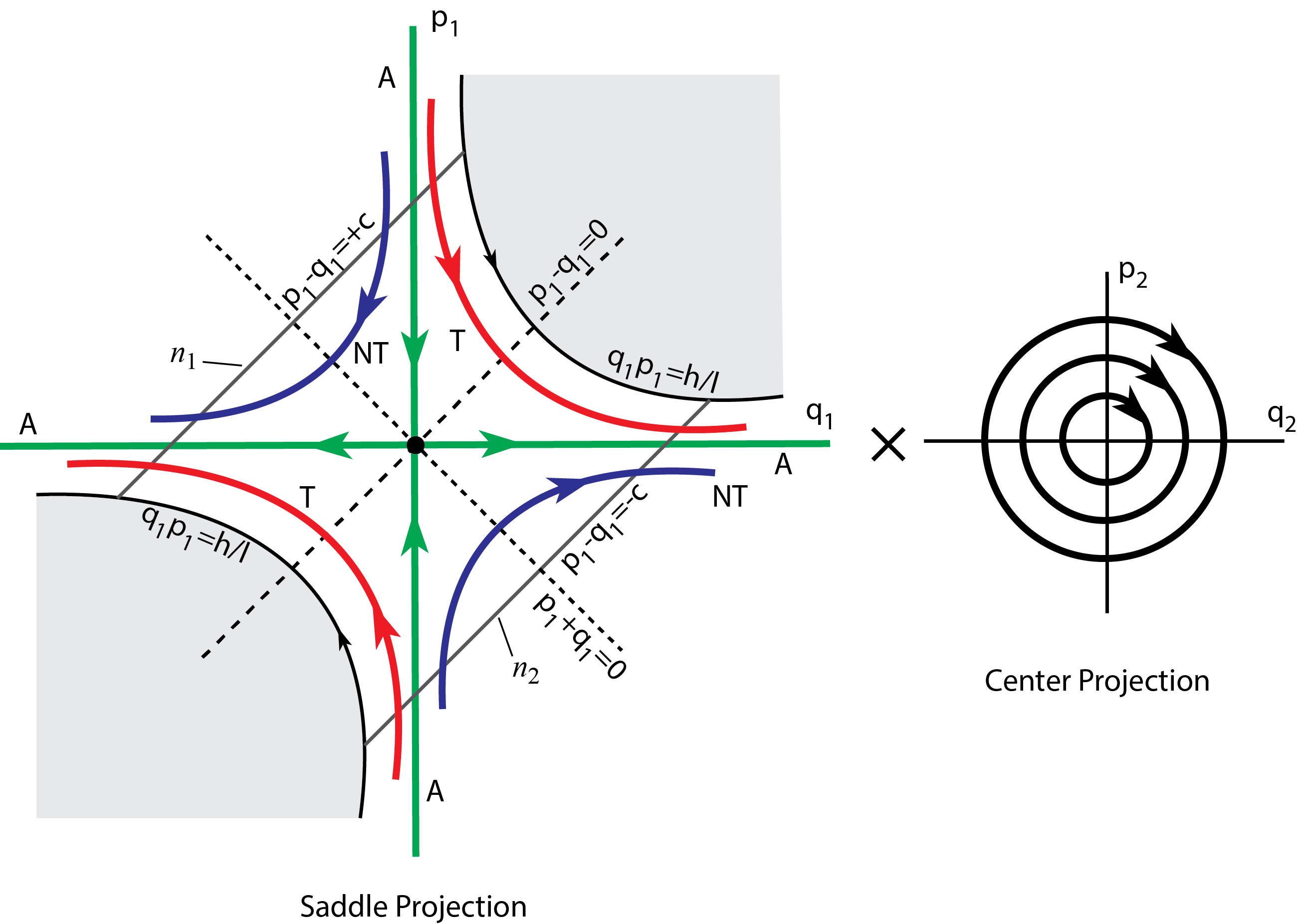}
\end{center}
\caption{\label{fig5}{\footnotesize 
The flow in the equilibrium region has the form
saddle $\times$ center.
On the left is shown the projection onto the $(p_1,q_1)$ plane, the saddle projection.
For the conservative dynamics, the Hamiltonian function $\mathcal{H}_2$ remains constant at $h>0$. Shown are the periodic orbit
(black dot at the center), the asymptotic orbits (labeled A), two
transit orbits (T) and two non-transit orbits (NT).
}}
\end{figure}

The bounding 2-sphere of $\mathcal{R}$ for which $p_1 -q_1 = c$ will
be called $n_1$ (the ``left'' bounding 2-sphere), and  that where $p_1 -q_1 = -c$,
$n_2$ (the ``right'' bounding 2-sphere). See Figure \ref{fig5}.  

We call the set of points on each
bounding 2-sphere where $q_1 + p_1 = 0$ the equator, and the sets where
$q_1 + p_1 > 0$ or $q_1 + p_1 < 0$ will be called the northern and 
southern hemispheres, respectively.

\paragraph{The Linear Flow in $\mathcal{R}$} 
To analyze the flow in
$\mathcal{R}$,   consider
the projections on the ($q_1, p_1$)-plane and the $(q_2,p_2)$-plane, respectively.
In the first case we see the standard
picture of a saddle point in two dimensions,
and in the second, of a center consisting of 
harmonic oscillator motion.
Figure \ref{fig5} schematically illustrates the flow.
With regard to the first projection we
see that $\mathcal{R}$ itself projects
to a set bounded on two sides by
the hyperbola
$q_1p_1 = h/\lambda $
(corresponding to $q_2^2+p_2^2=0$, see \eqref{Enlin}) and on two
other sides by the line segments
$p_1-q_1= \pm c$, which correspond to the bounding 2-spheres, $n_1$ and $n_2$, respectively.

Since $q_1p_1$ is an integral
of the equations in $\mathcal{R}$,
the projections of
orbits in the $(q_1,p_1)$-plane
move on the branches of the corresponding
hyperbolas $q_1p_1 =$ constant,
except in the case $q_1p_1=0$, where $q_1 =0$ or $p_1 =0$.
If $q_1p_1 >0$, the branches connect
the bounding line segments $p_1 -q_1 =\pm c$ and if $q_1p_1 <0$, they
have both end points on the same segment.  A check of equation
\eqref{new equa} shows that the orbits move
as indicated by the arrows in Figure \ref{fig5}.

To interpret Figure \ref{fig5} as a flow in $\mathcal{R}$,  notice
that each point in the $(q_1,p_1)$-plane projection 
corresponds to a 1-sphere $S^1$ in
$\mathcal{R}$ given by  
\[
q_2^2+p_2^2
=\tfrac{2 }{\omega_p}(h-\lambda q_1p_1) .
\]
Of course, for points on the bounding  hyperbolic
segments ($q_1p_1 =h/\lambda $), the
1-sphere collapses to a point. Thus, the segments of the lines 
$p_1-q_1 =\pm c$  in the projection correspond to the 2-spheres
bounding $\mathcal{R}$.  This is because each corresponds to a
1-sphere crossed with an interval where the two end 1-spheres are
pinched to a point.

We distinguish nine classes of orbits grouped into the following four
categories:
\begin{enumerate}
\item The point $q_1 =p_1 =0$ corresponds to an invariant
1-sphere $S^1_h$, an unstable {\bf period orbit} in
$\mathcal{R}$.  This 1-sphere is given by
\begin{equation}\label{3-sphere}
q_2^2+p_2^2=\tfrac{2 }{\omega_p}h, 
\hspace{0.3in} q_1 =p_1 =0.
\end{equation}
It is an example of a 
normally hyperbolic invariant manifold (NHIM) (see \cite{Wiggins1994}).
Roughly, this means that the stretching and contraction rates under
the linearized dynamics transverse to the 1-sphere dominate those tangent
to the 1-sphere.  This is clear for this example since the dynamics normal
to the 1-sphere are described by the exponential contraction and expansion
of the saddle point dynamics.  Here the 1-sphere acts as a ``big
saddle point''.  
See the black dot at the center of the $(q_1,p_1)$-plane on the left side 
of Figure
\ref{fig5}.
\item The four half open segments on the axes, $q_1p_1 =0$, 
correspond to four 
cylinders of orbits asymptotic to this invariant 1-sphere 
$S^1_h$ either as time
increases ($p_1 =0$) or as time decreases ($q_1 =0$).  These are called {\bf
asymptotic} orbits and they form the stable and the unstable manifolds of
$S^1_h$.  The stable manifolds, $W^s_{\pm}(S^1_h)$, are given by
\begin{equation}\label{stable_manifold}
q_2^2+p_2^2=\tfrac{2 }{\omega_p}h, 
 \hspace{0.3in} q_1 =0,
  \hspace{0.3in} p_1 ~{\rm arbitrary}.
\end{equation}
$W^s_+(S^1_h)$ (with $p_1>0$) is the branch going entering from $n_1$ and
$W^s_-(S^1_h)$ (with $p_1<0$) is the branch going entering from $n_2$.
The unstable manifolds, $W^u_{\pm}(S^1_h)$, 
are given by
\begin{equation}\label{unstable_manifold}
q_2^2+p_2^2=\tfrac{2 }{\omega_p}h, 
\hspace{0.3in} p_1 =0,
  \hspace{0.3in} q_1 ~{\rm arbitrary}
\end{equation}
$W^u_+(S^1_h)$ (with $q_1>0$) is the branch exiting from $n_2$ and
$W^u_-(S^1_h)$ (with $q_1<0$) is the branch exiting from $n_1$.
See the four orbits labeled A of Figure \ref{fig5}.
\item The hyperbolic segments determined by
$q_1p_1 ={\rm constant}>0$ correspond
to two cylinders of orbits 
which cross $\mathcal{R}$ from one bounding 2-sphere to the
other, meeting both in the same hemisphere; the northern hemisphere
if they go from
$p_1-q_1 =+c$ to $p_1-q_1 =-c$, and the southern hemisphere
in the other case. Since
these orbits transit from one realm to another, we call  them {\bf transit}
orbits.  See the two orbits labeled T of Figure \ref{fig5}.

\item Finally the hyperbolic segments determined by $q_1p_1 = {\rm
constant}<0$ correspond to two cylinders of orbits in
$\mathcal{R}$ each of which runs from one hemisphere to the other hemisphere
on the same bounding 2-sphere.  Thus if $q_1 >0$, the 2-sphere is $n_1$ ($p_1
-q_1 =-c$) and orbits run from the southern hemisphere
($q_1 +p_1 <0$) to the northern hemisphere ($q_1
+p_1 >0$) while the converse holds if $q_1 <0$, where the 
2-sphere is
$n_2$. Since these orbits return to the same realm, we call them {\bf
non-transit} orbits.  See the two orbits labeled NT of Figure \ref{fig5}.
\end{enumerate}

\subsection{Trajectories in the neck region}

We now examine the appearance of the orbits in configuration space, that is, in $(\bar q_1,\bar q_2)$-plane. In configuration space, $\mathcal{R}$ appears as the neck region connecting two realms, so trajectories in $\mathcal{R}$ will be transformed back to the neck region. It should pointed out that at each moment in time, all trajectories must evolve within the energy boundaries which are zero velocity curves (corresponding to $ \bar p_1 = \bar p_2 = 0$) given by solving \eqref{H_2_bar} for $\bar q_2$ as a function of $\bar q_1$,
\[
\bar q_2( \bar q_1) = \frac{\bar q_1 \pm \sqrt{\bar q_1^2 - 2 c_y (h + \tfrac{c_x}{2} \bar q_1^2)}}{c_y}
\]

Recall that in order to obtain the analytical solutions for $\bar z = (\bar q_1, \bar q_2, \bar p_1, \bar p_2)$, system $\bar z$ has been transformed into system $z = (q_1, q_2, p_1, p_2)$ by using the symplectic matrix $C$ consisting of generalized  (re-scaled) eigenvectors $u_{+ \lambda}, u_{- \lambda}, u_{\omega_p}, v_{\omega_p}$ with corresponding eigenvalues $ \pm \lambda$ and $\pm i \omega_p$. Thus, the system $z$ should be transformed back to system $\bar z$ which  generates the following general (real) solution with the form
 \begin{equation}
 \begin{split}
 \bar z(t) = \left(\bar q_1, \bar q_2, \bar p_1, \bar p_2 \right)^T = q_1^0 e^{\lambda t} u_{+ \lambda} + p_1^0 e^{- \lambda t} u_{- \lambda} + \mathrm{Re} \left[\beta_0 e^{- i \omega_p t}    \left(u_{\omega_p} - i v_{\omega_p} \right) \right]
 \label{whole-gener-sol}
 \end{split}
 \end{equation}
where $q_1^0, p_1^0$ are real and $\beta_0 = q_2^0 + i p_2^0$ is complex.

\begin{figure}[h!]
	\begin{center}
		\includegraphics[width=0.67\textwidth]{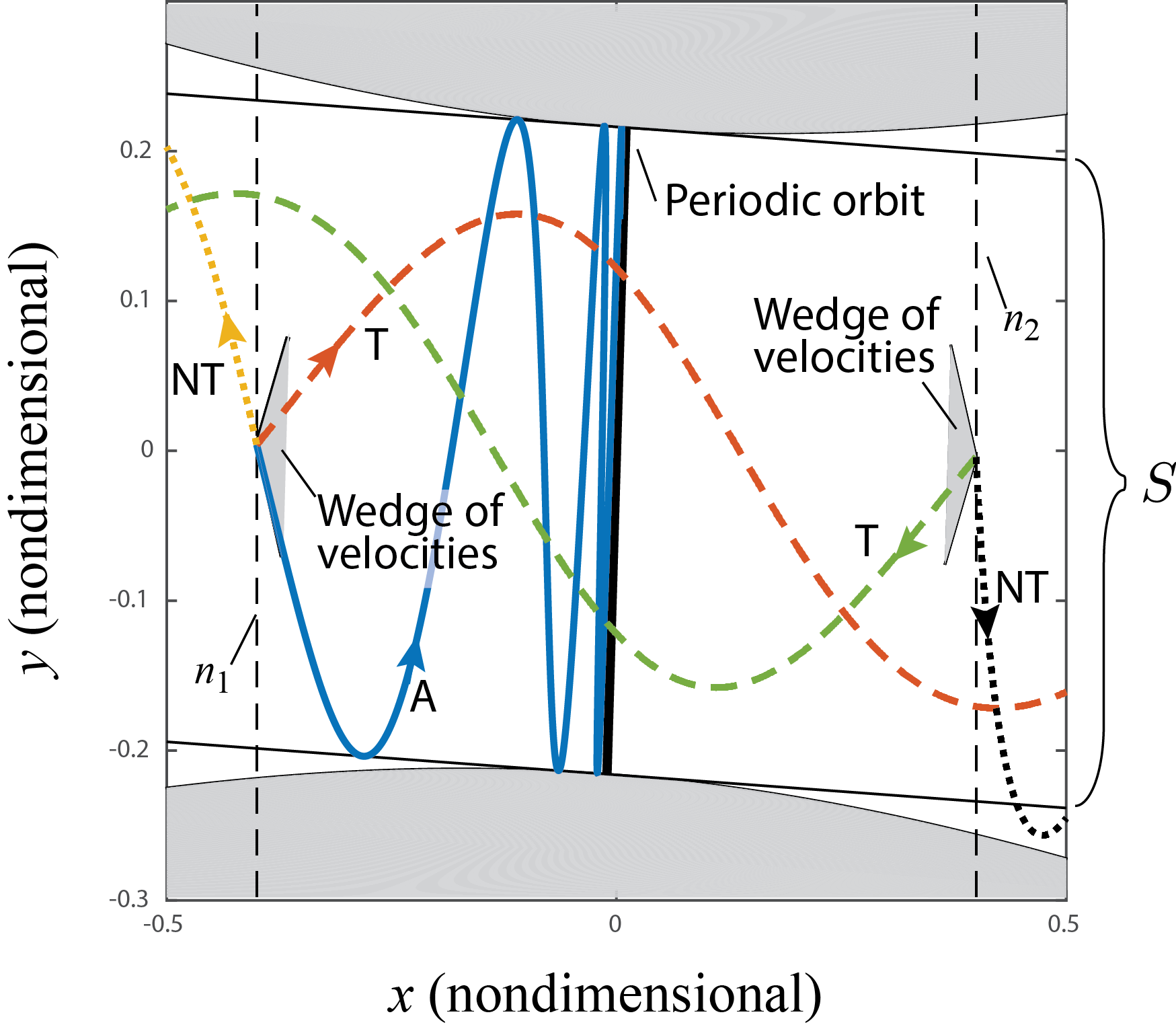}
		\caption{
The flow of the conservative system in  $\mathcal{R}$, the equilibrium region projected onto the $xy$ configuration space, for a fixed value of energy, $\mathcal{H}_2=h>0$.
For any point on the bounding vertical lines $n_1$ or $n_2$ (dashed), there is a wedge of velocity directions inside of which the trajectories are transit orbits, and outside of which are non-transit orbits. The boundary of the wedge gives the orbits asymptotic to the single unstable periodic orbit in the neck for this energy.
Shown are a typical asymptotic orbit; two transit orbits (dashed); and two non-transit orbits (dotted).
		}
		\label{fig:Conley}
	\end{center}
\end{figure}

Upon inspecting this general solution, we see that the solutions on the energy surface fall into different classes depending upon the limiting behaviors of $ \bar q_1, \bar q_2$ as $t$ tends to plus or minus infinity. Notice that
\begin{equation}
       \begin{split}
              \bar q_1(t) &= \frac{q_1^0}{s_1} e^{\lambda t} - \frac{p_1^0}{s_1} e^{-\lambda t} + \frac{1}{s_2} \left( q_2^0 \cos \omega_p t + p_2^0 \sin \omega_p t \right)\\
              \bar q_2(t) &= \frac{c_x-\lambda^2}{s_1} q_1^0 e^{\lambda t} + \frac{\lambda^2 - c_x}{s_1} p_1^0 e^{-\lambda t} + \frac{\omega_p^2 +c_x}{s_2} \left( q_2^0 \cos \omega_p t + p_2^0 \sin \omega_p t \right)\\
              \label{conser-sol}
       \end{split}
\end{equation}
Thus, if $t\rightarrow + \infty$, then $\bar q_1 (t)$ is dominated by its $q_1^0$ term. Hence,  $\bar q_1 (t)$ tends to minus infinity (staying on the left-hand side), is bounded (staying around the equilibrium point), or tends to plus infinity (staying on the right-hand side) according to $q_1^0 < 0 $, $q_1^0=0$ and $q_1^0>0$. See Figure \ref{fig:Conley}. The same statement holds if $t\rightarrow - \infty$ and $-p_1^0$ replaces $q_1^0$. Different combinations of the signs of $q_1^0$ and $p_1^0$ will give us again the same nine classes of orbits which can be grouped into the same four categories.

\begin{enumerate}
	\item If $q_1^0 = p_1^0 = 0$, we obtain a periodic solution. 
	The periodic orbit projects onto the $(\bar q_1, \bar q_2)$ plane as a line with the following expression
	\begin{equation}
	\begin{split}
	\bar q_1 & = \frac{1}{s_2} \left( q_2^0 \cos \omega_p t + p_2^0 \sin \omega_p t \right)\\
	\bar q_2 & = \frac{ \omega_p^2 + c_x}{s_2} \left(q_2^0 \cos \omega_p t + p_2^0 \sin \omega_p t \right)\\
	& =\left( \omega_p^2 + c_x \right) \bar q_1
	\end{split}
	\end{equation} 
	Notice \eqref{Enlin} and $\mathcal{H}_2$ now can be rewritten as $\mathcal{H}_2 = \omega_p |\beta_0|^2 / 2$. Thus, since $\mathcal{H}_2=h$, the length of the periodic orbit is  $\sqrt{  2 h \left[(\omega_p^2 + c_x)^2+1 \right] / \left(\omega_p s_2^2 \right)}$. Note that the length of the line goes to zero with $h$.
	
	\item Orbits with $q_1^0 p_1^0=0$ are asymptotic orbits. They are asymptotic to the periodic orbit.
	\begin{enumerate}
		\item When $q_1^0=0$ , the general solutions for $\bar q_1, \bar q_2$ are
		\begin{equation}
		\begin{split}
		\bar q_1 &= - \frac{p_1}{s_1}+\frac{q_2}{s_2} \\
		\bar q_2 &= \frac{ \lambda^2 - c_x}{s_1} p_1 + \frac{\omega_p^2 + c_x}{s_2} q_2 \\
		& = \left(c_x - \lambda^2 \right) \bar q_1 + \frac{\lambda^2 + \omega_p^2}{s_2} \left(q_2^0 \cos \omega_p t + p_2^0 \sin \omega_p t \right)
		\end{split}		           
		\end{equation}
		Thus, the orbits with $q_1^0=0$ project into a strip $S$ in the $(\bar q_1, \bar q_2)$-plane bounded by
		\begin{equation}
		       \begin{split}
		           \bar q_2 =\left(c_x - \lambda^2 \right) \bar q_1 \pm \frac{\lambda^2 + \omega_p^2}{s_2} \sqrt{\frac{2 h}{\omega_p}}
		           \label{strip boundary}
		       \end{split}
		\end{equation}

		\item For $p_1^0=0$, following the same procedure as $q_1^0=0$, we have
		\begin{equation}
		\begin{split}
		\bar q_1 &= \frac{q_1}{s_1} + \frac{q_2}{s_2}\\
		\bar q_2 &= \left(c_x - \lambda^2\right) \bar q_1 + \frac{\lambda^2 + \omega_p^2}{s_2} \left(q_2^0 \cos \omega_p t + p_2^0 \sin \omega_p t \right)
		\end{split}
		\end{equation}
		Notice that these two asymptotic orbits with $q_1^0=0$ and $p_1^0=0$ share the same strip $S$ and the same boundaries governed by \eqref{strip boundary}. Also, since the slopes of the periodic orbit and the strip satisfies $\left(c_x - \lambda^2 \right) \left(c_x + \omega_p^2 \right)=-1$, the periodic orbit is perpendicular to the strip. In other words, the length of the periodic orbit is exactly the same as the width of the strip.
	\end{enumerate}
	
	\item Orbits with $q_1^0 p_1^0 >0$ are transit orbits because they cross the equilibrium region $R$ from $- \infty$ (the left-hand side) to $+ \infty$ (the right-hand side) or vice versa.
	
	\item Orbits with $q_1^0 p_1^0 <0$ are non-transit orbits
\end{enumerate}

To study the flow in position space, Figure \ref{fig:Conley} gives the four categories of orbits. From  \eqref{conser-sol}, we can see that for transit orbits and non-transit orbits, the signs of $q_1^0 p_1^0$ must satisfy $q_1^0 p_1^0>0$ and $q_1^0 p_1^0<0$,respectively. 
 
In Figure \ref{fig:Conley}, $S$ is the strip mentioned above. Outside of the strip, the signs of $q_1^0$ and $p_1^0$ are independent of the direction of the velocity. These signs can be determined in each of the components of the equilibrium region $\mathcal{R}$ complementary to the strip. For example, in the left two components, $q_1^0<0$ and $p_1^0>0$, while in the right two components $q_1^0>0$ and $p_1^0<0$. Therefore, $q_1^0 p_1^0<0$ in all components and only non-transit orbits project on to these four components.
 
Inside the strip the situation is more complicated since in $S$ the signs of $q_1^0$ and $p_1^0$ depend on the direction of the velocity. At each position $(\bar q_1, \bar q_2)$ inside the strip there exists the so-called `wedge' of velocities in which $q_1^0 p_1^0>0$ which was first found by Conley (1968) \cite{Conley1968} in the restricted three-body problem. See the shaded wedges in Figure \ref{fig:Conley}. The existence and the angle of the wedge of velocity will be given in the next part. For simplicity we have indicated this dependence only on the two vertical bounding line segments in Figure \ref{fig:Conley}. For example, consider the intersection of strip $S$ with left-most vertical line. On the subsegment so obtained there is at each point a wedge of velocity in which both $q_1^0$ and $p_1^0$ are positive, so that orbits with velocity interior to the wedge are transit orbits $(q_1^0 p_1^0>0)$. Of course, orbits with velocity on the boundary ot the wedge are asymptotic $(q_1^0 p_1^0 = 0)$, while orbits with velocity outside of the wedge are non-transit. The situation on the other subsegment is similar. 

\paragraph{The wedge of velocities} 
To establish the wedge of velocity and obtain its angle, we need to use the following fact that all the inner products of one generalized eigenvector and another generalized eigenvector associated with 
$B$ are zero except for 
\begin{equation}
       \begin{split}
              u_{+ \lambda}^T B u_{- \lambda} &=u_{- \lambda}^T B u_{+ \lambda} = \lambda \\
              u_{\omega_p}^T B u_{\omega_p} &= v_{\omega_p}^T B v_{\omega_p} = \omega_p \\
       \end{split}
\end{equation}

Using this condition, we have the following relations, as
 \begin{equation}
        \begin{split}
                \lambda &= u_{+ \lambda}^T B u_{- \lambda}\\
               \Rightarrow  \lambda q_1^0 &= q_1^0 u_{+ \lambda}^T B u_{- \lambda}\\
               \Rightarrow  \lambda q_1^0 &= e^{- \lambda t} \left(q_1^0 e^{\lambda t} u_{+ \lambda} \right)^T B u_{- \lambda}\\
               \Rightarrow  \lambda q_1^0 &= e^{- \lambda t} \bar z^T B u_{- \lambda}\\
               \Rightarrow \lambda q_1^0 &= e^{- \lambda t} \left(\frac{\lambda^2}{s_1} \bar q_1 - \frac{ 1-c_xc_y + c_y \lambda^2}{s_1} \bar q_2 + \frac{\lambda}{s_1} \bar p_1 + \frac{c_x \lambda - \lambda^3}{s_1} \bar p_2 \right)
        \end{split}
 \end{equation}
Using similar arguments, we can also obtain
\begin{equation}
       \begin{split}
              \lambda p_1^0 &= e^{\lambda t} \left(- \frac{\lambda^2}{s_1} \bar q_1 + \frac{1- c_x c_y + c_y \lambda^2}{s_1} \bar q_2 + \frac{\lambda}{s_1} \bar p_1 + \frac{c_x \lambda - \lambda^3}{s_1} \bar p_2 \right)
       \end{split}
\end{equation}
Thus, we obtain the following relations
\begin{equation}
       \begin{split}
            \lambda q_1^0 e^{\lambda t}&=  \frac{\lambda^2}{s_1} \bar q_1 - \frac{1- c_x c_y + c_y \lambda^2}{s_1} \bar q_2 + \frac{\lambda}{s_1} \bar p_1 + \frac{c_x \lambda - \lambda^3}{s_1} \bar p_2\\
            \lambda p_1^0 e^{- \lambda t} &=- \frac{\lambda^2}{s_1} \bar q_1 + \frac{1- c_x c_y + c_y \lambda^2}{s_1} \bar q_2 + \frac{\lambda}{s_1} \bar p_1 + \frac{c_x \lambda - \lambda^3}{s_1} \bar p_2
            \label{chi prepare}
       \end{split}
\end{equation}

Let $\chi$  be the angles determined by
\begin{equation}
       \begin{split}
             \cos \chi &= \frac{1}{\sqrt{ \left(\lambda^2 -c_x \right)^2 + 1}}, \ \ \ \ \ \ \sin \chi = \frac{\lambda^2 -c_x}{\sqrt{ \left(\lambda^2 -c_x \right)^2 + 1}}.\\ 
       \end{split}
\end{equation}
Furthermore, let
\begin{equation}
       \begin{split}
             \bar p_1 = \rho \cos \theta, \ \ \ \bar p_2= \rho \sin \theta
       \end{split}
\end{equation}
and
\begin{equation}
       \begin{split}
              \gamma &= \left( \frac{\lambda^2}{s_1} \bar q_1 - \frac{1- c_x c_y + c_y \lambda^2}{s_1} \bar q_2 \right) \left[\frac{\lambda^2}{s_1^2} \left(\bar p_1^2 + \bar p_2 ^2\right) \left(\left(\lambda^2 -c_x \right)^2 + 1 \right) \right]^{- \frac{1}{2}}\\
              \label{gamma}
       \end{split}
\end{equation}
Using \eqref{gamma}, \eqref{chi prepare} can be rewritten as
\begin{equation}
       \begin{split}
             \lambda q_1^0 e^{\lambda t} \left[\frac{\lambda^2}{s_1^2} \left(\bar p_1^2 + \bar p_2 ^2\right) \left(\left(\lambda^2 -c_x \right)^2 + 1 \right) \right]^{- \frac{1}{2}} &= \gamma + \cos(\theta - \chi)\\
             \lambda p_1^0 e^{- \lambda t} \left[\frac{\lambda^2}{s_1^2} \left(\bar p_1^2 + \bar p_2 ^2\right) \left(\left(\lambda^2 -c_x \right)^2 + 1 \right) \right]^{- \frac{1}{2}} &= - \gamma + \cos(\theta - \chi)
             \label{angle of wedge}
       \end{split}
\end{equation}

So far, the signs of $q_1^0$ and $p_1^0$ can be determined using Eq. \eqref{angle of wedge}. From Eq. \eqref{angle of wedge}, it can be concluded  that $\gamma$ is only dependent on the position $(\bar q_1, \bar q_2)$, because $\bar p_1^2 + \bar p_2 ^2$ can be obtained from Eq. \eqref{H_2_bar} once the position is given. Outside the strip, we have $\mid \gamma \mid >1 $. In this case, the signs of $q_1^0$ and $p_1^0$ are independent of the direction of velocity and are always opposite, which makes $q_1^0 p_1^0<0$. Thus, only non-transit orbit exist in these regions. Inside the strip, we have $\mid \gamma \mid <1 $. This situation is quite different since the signs of $q_1^0$ and $p_1^0$ are dependent on the angle of velocity. Because for transit orbits, the sign of $q_1^0 p_1^0$ must be positive.  Thus, we can vary $\theta$ (the direction of velocity) to satisfy this condition, and the wedge of velocity can be determined. It should be noted that the wedge of velocity can only exist inside the strip $S$: outside of $S$, no transit orbit exists.

\section{Linearized Dissipative Hamiltonian System} \label{linearization of dissipative}

\subsection{Solutions near the equilibria}

For the dissipative system, we still use the symplectic matrix $C$ as in \eqref{sym matrix}  to transform to the eigenbasis, i.e., transform $\bar z=(\bar q_1 , \bar q_2 , \bar p_1 , \bar p_2)$ to $z=(q_1 , q_2 , p_1 , p_2)$. The equations of motion now become
\begin{equation}
	\dot z = \Lambda  z + \Delta z
\end{equation}
where $\Lambda  = {C}^{-1}{ {A}}{C}$ from before and the transformed damping matrix is,
\begin{equation}
	\begin{split}
		\Delta = { C}^{-1}{{D}}{C} = -c_1 \left( \begin{array}{rrrr}
			\tfrac{1}{2} 	&  0 & \tfrac{1}{2}  	& 0 \\
			0 		& 0 & 0	 		& 0 \\
			\tfrac{1}{2} 	&  0 & \tfrac{1}{2}  	& 0 \\
			0	  	& 0 & 0 			& 1 
		\end{array} \right)
		\label{zeta_matrix}
	\end{split}
\end{equation}
which results in
\begin{subequations}	
	\begin{align}
		&\begin{cases}
			\dot q_1 = \left(\lambda - \frac{ c_1}{2}   \right) q_1 - \frac{ c_1}{2} p_1\\
			\dot p_1 = - \frac{ c_1}{2} q_1 + \left(- \lambda -\frac{ c_1 }{2}  \right) p_1	
			\label{Hd q1}
		\end{cases}\\
		&\begin{cases}
			\dot q_2 = \omega_p p_2\\
			\dot p_2 = - \omega_p q_2 - c_1 p_2
			\label{Hd q2}
		\end{cases}		
	\end{align}	
\end{subequations}
Notice that the dynamics on the $(q_1,p_1)$ plane and $(q_2,p_2)$ plane are uncoupled.

The fourth-order characteristic polynomial is thus decomposable into $p(\beta)= p_1(\beta)  p_2(\beta)$, where
the second-order characteristic polynomials for \eqref{Hd q1} and \eqref{Hd q2} are 
\begin{subequations}
	\begin{numcases}{}
		p_1(\beta)= \beta^2 + c_1 \beta - \lambda^2\\
		p_2(\beta)= \beta^2 + c_1 \beta + \omega_p^2
	\end{numcases}
	\label{Hd poly}
\end{subequations}
Considering $c_1$ is positive and $c_1^2$ is smaller compared with $4 \omega_p^2$, the determinants for \eqref{Hd poly} are
\begin{subequations}
	\begin{numcases}{}
		\Delta_1 = c_1^2 + 4 \lambda^2 > 0\\
		\Delta_2 = c_1^2 - 4 \omega_p^2 <0
	\end{numcases}
\end{subequations}
The corresponding eigenvalues are
\begin{subequations}	
	\begin{align}
		&\begin{cases}
			\beta_1= \frac{ - c_1 + \sqrt{ c_1^2 + 4 \lambda^2}}{2}\\
			\beta_2= \frac{ - c_1 - \sqrt{ c_1^2 + 4 \lambda^2}}{2}	
		\end{cases}\\
		&\begin{cases}
			\beta_3= -\delta + i \omega_d \\
			\beta_4= -\delta  - i \omega_d \\
		\end{cases}		
	\end{align}	
\end{subequations}
where $\delta = \frac{c_1}{2}, \omega_d =\omega_p  \sqrt{ 1 - \xi_d^2}$ and $\xi_d=\frac{\delta}{\omega_p}$, with the corresponding eigenvectors
\begin{equation}
	\begin{split}
		u_{\beta_1}&= \left( \frac{c_1}{2}, \lambda - \frac{1}{2} \sqrt{c_1^2 + 4 \lambda^2} \right)\\
		u_{\beta_2}&= \left( \frac{c_1}{2}, \lambda + \frac{1}{2} \sqrt{c_1^2 + 4 \lambda^2} \right)\\
		u_{\beta_3}&= \left( \omega_p, - \delta + i \omega_d \right)\\
		u_{\beta_4}&= \left( \omega_p, - \delta - i \omega_d \right)
		\label{Hd imagin}
	\end{split}
\end{equation}

Thus, the general solutions for the $\left(q_1,p_1\right)$ and $\left(q_2,p_2\right)$ systems are 
\begin{subequations}	
	\begin{align}
		&\begin{cases}
			q_1= k_1 e^{\beta_1 t} + k_2 e^{\beta_2 t}\\
			p_1= k_3 e^{\beta_1 t} + k_4 e^{\beta_2 t}
		\end{cases}\\
		&\begin{cases}
			q_2= k_5  e^{- \delta t} \cos{\omega_d t} + k_6 e^{- \delta t} \sin{\omega_d t}\\
			p_2= \frac{k_5 }{\omega_p} e^{- \delta t} \left(-\delta \cos{\omega_d t} - \omega_d \sin{\omega_d t} \right) +\frac{k_6 }{\omega_p}  e^{- \delta t} \left(\omega_d \cos{\omega_d t - \delta \sin{\omega_d t}} \right)\\
		\end{cases}		
	\end{align}
	\label{Hd gener solut}
\end{subequations}
where
\begin{equation*}
	\begin{split}
		k_1 &= \frac{q_1^0 \left(2 \lambda + \sqrt{c_1^2 + 4 \lambda^2} \right)-c_1 p_1^0 }{2\sqrt{c_1^2 + 4 \lambda^2}}, \hspace{0.5in} k_2 =\frac{q_1^0 \left(-2 \lambda + \sqrt{c_1^2 + 4 \lambda^2} \right)+c_1 p_1^0 }{2\sqrt{c_1^2 + 4 \lambda^2}},\\
		k_3 &= \frac{p_1^0 \left(-2 \lambda + \sqrt{c_1^2 + 4 \lambda^2} \right)-c_1 q_1^0 }{2\sqrt{c_1^2 + 4 \lambda^2}}, \hspace{0.5in} k_4 = \frac{p_1^0 \left(2 \lambda + \sqrt{c_1^2 + 4 \lambda^2} \right)+c_1 q_1^0 }{2\sqrt{c_1^2 + 4 \lambda^2}},\\
		k_5&=q^0_2 , \hspace{0.5in} k_6=\frac{p^0_2 \omega_p + q^0_2 \delta}{\omega_d} 
	\end{split}
\end{equation*}
Note that $k_1=q_1^0$, $k_4=p_1^0$, $k_2=k_3=0$, $k_5=q_2^0$ and $k_6=p_2^0 $ if $c_1=c_2=0$.

Taking total derivative with respect to $t$ of the Hamiltonian along trajectories gives us
\begin{equation}
	\begin{split}
		\frac{d \mathcal{H}_2}{d t} = - \tfrac{1}{2} c_1 \lambda \left(q_1 + p_1 \right)^2 - c_1 \omega_p p_2^2 \le 0
		\label{H2-rate of change-equal}
	\end{split}
\end{equation}
which means the Hamiltonian is non-increasing, and will generally decrease due to damping.

\subsection{Boundary of transit and non-transit orbits}

\begin{figure}[!htb]
	\begin{center}
		\includegraphics[width=\textwidth]{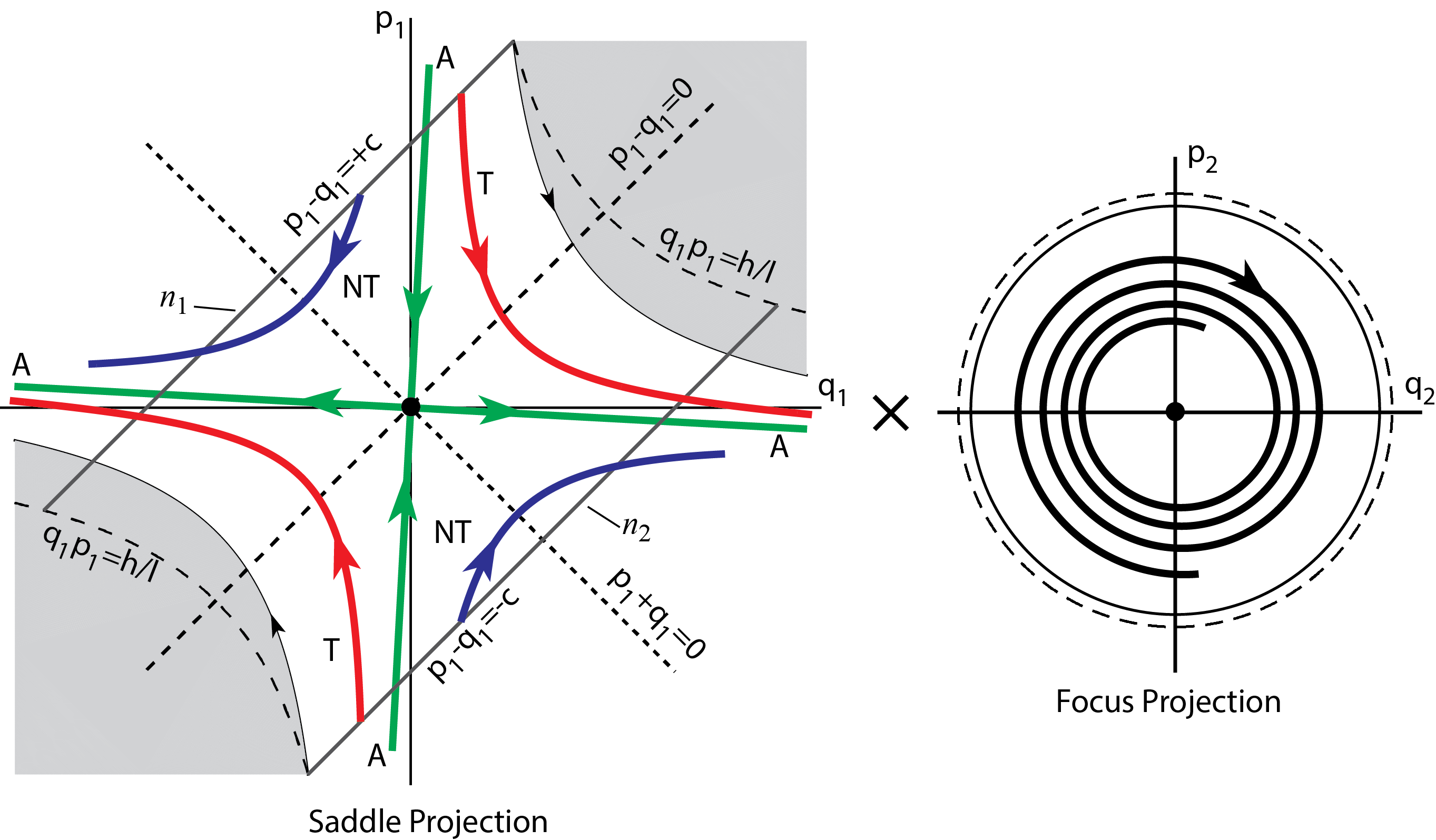} 			
	\end{center}
	\caption{\label{flow-damped}{\footnotesize
	The flow in the equilibrium region around S$_1$ for the dissipative system has the form
saddle $\times$ focus.
On the left is shown the projection onto the $(p_1,q_1)$ plane, the saddle projection. The
asymptotic orbits (labeled A) on this projection are the saddle-type asymptotic orbits, and are rotated clockwise compared to the conservative system.
They still form the separatrix between transit orbits (T) and two non-transit orbits (NT).
The black dot at the center  represents trajectories with only a focus projection, thus oscillatory dynamics decaying onto the point S$_1$.
As the energy, the Hamiltonian function $\mathcal{H}_2$, is decreasing, the boundary is no longer equal to $q_1 p_1 = h/\lambda$, as it is for the conservative case, where $\mathcal{H}_2=h$ is the initial value of the energy for those trajectories entering through the left or right side bounding sphere (i.e., $n_1$ or $n_2$, respectively). These boundaries (the boundary of the shaded region) still correspond to the fastest trajectories through the neck region for a given $h$.
			}}
\end{figure}

\paragraph{The Linear Flow in $\mathcal{R}$} 
Similar to the discussions for the conservative system, we still choose an equilibrium region $\mathcal{R}$ bounded by regions which project to the lines $n_1$ and $n_2$ in the  $(q_1,p_1)$-plane (see Figure \ref{flow-damped}).
To analyze the flow in $\mathcal{R}$, we consider the projections on the $(q_1,p_1)$-plane and the $(q_2,p_2)$-plane, respectively. In the first case we see the standard picture of saddle point, now rotated compared to the conservative case, and in the second, of a stable focus which is a damped oscillator with frequency $\omega_d=\omega_p \sqrt{1- \xi_d^2}$, where $\xi_d=\frac{c_1}{2 \omega_p}$ - the viscous damping factor (damping ralative to critical damping). Notice that the frequency $\omega_d$ for the damped system decreases with increased damping, but only very slightly for lightly damped systems.

We distinguish nine classes of orbits grouped into the following four categories:
\begin{enumerate}
	\item The point $q_1=p_1=0$ corresponds to a  {\bf focus-type asymptotic} orbit with motion purely in the $(q_2,p_2)$-plane (see black dot at the origin of
	the $(q_1,p_1)$-plane in Figure \ref{flow-damped}).  
	Such orbits are asymptotic to the equilibrium point S$_1$ itself.
	Due to the effect of damping, the periodic orbit in the conservative system, which is an invariant 1-sphere $S_h^1$ mentioned in \eqref{3-sphere}, does not exist.   
	
	\item The four half open segments on the lines governed by $q_1=  c_1 p_1/(2 \lambda \pm \sqrt{c_1^2 + 4 \lambda^2}) $ correspond to {\bf saddle-type asymptotic} orbits.
	See the four orbits labeled A in Figure \ref{flow-damped}. These orbits have motion in both the  $(q_1,p_1)$- and $(q_2,p_2)$-planes.
	
	\item The segments which cross $\mathcal{R}$ from one boundary to the other, i.e., from $p_1 - q_1=+c$ to $p_1 - q_1=-c$ in the northern hemisphere, and vice versa in the southern hemisphere, correspond to {\bf transit} orbits. See the two orbits labeled $T$ of Figure \ref{flow-damped}.
	
	\item  Finally the segments which run from one hemisphere to the other hemisphere on the same boundary, namely which start from $p_1 - q_1 = \pm c$ and return to the same boundary, correspond to {\bf non-transit} orbits. See the two orbits labeled NT of Figure \ref{flow-damped}.
\end{enumerate}

\subsection{Trajectories in the neck region}

Following the same procedure of analysis as for conservative system, the general solution to the dissipative system can be obtained by $\bar z=C z$ which gives 
\begin{equation}
	\begin{split}
		\bar q_1 &= \frac{k_1 - k_3}{s_1} e^{\beta_1 t} - \frac{k_4 - k_2}{s_1} e^{\beta_2 t} + \frac{q_2}{s_2}\\
		\bar q_2 &= \frac{k_1 - k_3}{s_1} (c_x-\lambda^2) e^{\beta_1 t} - \frac{k_4 - k_2}{s_1} (c_x-\lambda^2) e^{\beta_2 t} + \frac{\omega_p^2 + c_x}{s_2} q_2
		\label{diss sol}
	\end{split}
\end{equation}

\begin{figure}[!htb]
	\begin{center}
			\includegraphics[width=\textwidth]{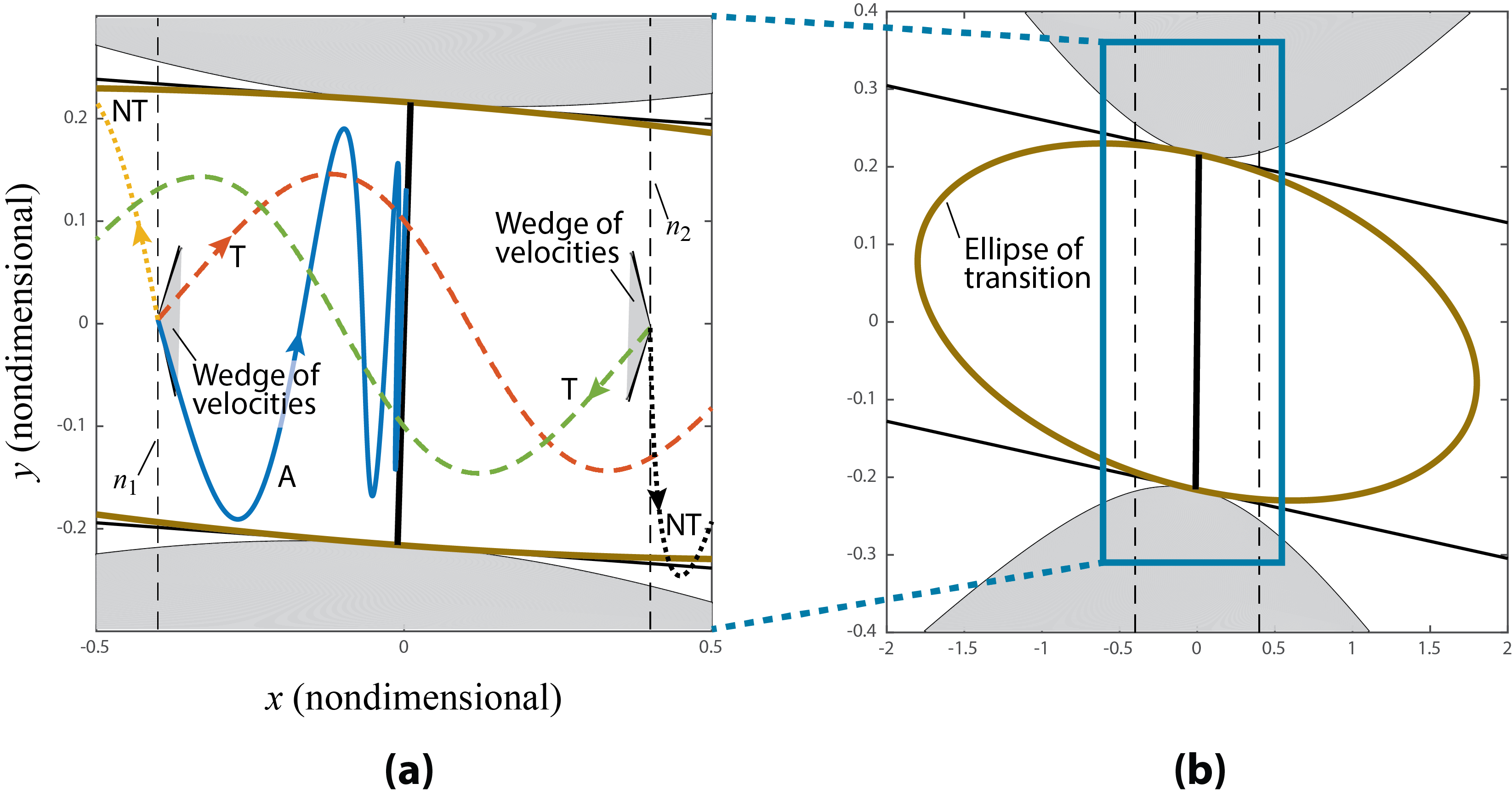} 
			\caption{\label{lin_damp_position_paper}{\footnotesize 
The flow of the dissipative system in  $\mathcal{R}$, the equilibrium region projected onto the $xy$ configuration space, for trajectories starting at a fixed value of energy, $\mathcal{H}_2=h$, on either the right or left side vertical boundaries.
As before, for any point on a bounding vertical line (dashed), there is a wedge of velocities inside of which the trajectories are transit orbits, and outside of which are non-transit orbits. For a given fixed energy, the wedge for the dissipative system is a subset of the wedge for the conservative system. The boundary of the wedge gives the orbits asymptotic (saddle-type) to the equilibrium point S$_1$.
}}
	\end{center}
\end{figure}

Similar to the situation in the conservative system, the solutions for the dissipative system on the energy surface fall into different classes depending upon the limiting behaviors.
See Figure \ref{lin_damp_position_paper}.
From  \eqref{diss sol} we know that the conditions $k_1-k_3>0$, $k_1 - k_3=0$ and $k_1-k_3<0$ make $\bar q_1$ tend to minus infinity, are bounded or tend to plus infinity if $t \rightarrow \infty$. See Figure \ref{flow-damped}. The same statement holds if $t \rightarrow- \infty$ and $k_2 -k_4$ replaces $k_1- k_3$. Nine classes of orbits can be given according to different combinations of the sign of $k_1-k_3$ and $k_2 - k_4$ which can be classified into the following four categories:
\begin{enumerate}
	\item Orbits with $k_1 - k_3 = k_4-k_2 = 0$ are {\bf focus-type asymptotic} orbits
	\begin{equation}
		\begin{split}
			\bar q_1 = q_2 / s_2, \hspace{0.5in} \bar q_2 = \left(\omega_p^2 + c_x \right) \bar q_1
		\end{split}
	\end{equation}
	Notice the presence of $q_2$ in \eqref{Hd gener solut} reveals that the amplitude of the periodic orbit will gradually decease at the rate of $e^{-\delta t}$ with time. The larger the damping, the faster the rate will be. 
	
	\item Orbits with $ \left(k_1 - k_3\right) \left(k_4 - k_2 \right)= 0$ are {\bf saddle-type asymptotic} orbits
	\begin{equation}
		\bar q_2 =\left(c_x-\lambda^2 \right) \bar q_1 + \frac{\lambda^2+\omega_p^2}{s_2} q_2
		\label{damped-sympt}
	\end{equation}
	In similarity with the shrinking of the length of the periodic orbit, the amplitude of asymptotic orbits are also shrinking.
	
	\item Orbits with $\left(k_1 - k_3\right) \left(k_4 - k_2 \right)>0$ are {\bf transit} orbits
	\item Orbits with $\left(k_1 - k_3\right) \left(k_4 - k_2 \right)<0$ are {\bf non-transit} orbits
\end{enumerate}

\paragraph{Wedge of velocities}
We previously obtained the wedge of velocities for the conservative system. However, this method is no longer effective for the dissipative system. Thus, another approach will be pursued here. 

Based on the eigenvectors in \eqref{Hd imagin}, we can conclude that the directions of stable asymptotic orbits are along $u_{\beta_2}=\left(\frac{c_1}{2}, \lambda + \frac{1}{2} \sqrt{c_1^2 + 4 \lambda^2} \right)$. In this case, all asymptotic orbits in the transformed system must start on the line
\begin{equation}
	q_1=k_p p_1
	\label{R:asymp-formula}
\end{equation}
where $k_p=c_1 / (2 \lambda + \sqrt{c_1^2 + 4 \lambda^2})$.

For a specific point $\left(\bar q_{10}, \bar q_{20} \right)$, the initial conditions in position space and transformed space are defined as $\left(\bar q_{10}, \bar q_{20}, \bar p_{10}, \bar p_{20} \right)$ and $\left(q_{10}, q_{20}, p_{10}, p_{20} \right)$, respectively. Using Eq. \eqref{R:asymp-formula} and the change of variables in \eqref{sym matrix}, we can obtain $p_{10}$, $q_{20}$,  $p_{20}$ and $\bar p_{20}$ in terms of $\bar q_{10}$, $\bar q_{20}$ and $\bar p_{10}$. With $p_{10}$, $q_{20}$,  $p_{20}$ and $\bar p_{20}$ in hand, the normal form of the Hamiltonian can be rewritten as
\begin{equation}
	a_p \bar p_{10}^2 + b_p \bar p_{10}+c_p=0
	\label{quadratic}
\end{equation}
where
\begin{equation*}
	\begin{split}
		& a_p=\frac{s_2^2}{2 \omega_p}, \hspace{0.5in} b_p=\frac{\lambda s_2^2 (1+k_p) \left[\bar q_2- \bar q_1 \left(c_x + \omega_p^2 \right) \right]}{\omega_p \left(k_p-1 \right) \left(\lambda^2 + \omega_p^2 \right)},\\
		& c_p = \left(\sum\limits_{i=1}^{4} c_p^{(i)}\right)/ \left[2 \omega_p \left(k_p-1 \right)^2 \left(\lambda^2 + \omega_p^2 \right)^2 \right]- h,\\
		& c_p^{(1)}=2 k_p s_1^2 \lambda \omega_p \left[\bar q_2- \bar q_1 \left(c_x + \omega_p^2 \right) \right]^2, \\
		& c_p^{(2)}= 8 k_p s_2^2 \lambda^2 \omega_p^2 \bar q_1 \left(c_x \bar q_1 - \bar q_2 \right),\\
		& c_p^{(3)}=s_2^2 \lambda^2 \left(1+k_p \right)^2 \left[\left(c_x \bar q_1- \bar q_2 \right)^2+ \bar q_1^2\omega_p^4 \right],\\
		& c_p^{(4)}=s_2^2 \omega_p^2 \left(k_p -1 \right)^2 \left[\left(c_x \bar q_1 - \bar q_2) \right)^2+ \bar q_1^2 \lambda^4 \right]
	\end{split}
\end{equation*}

For the existence of real solutions, the determinant of quadratic equation \eqref{quadratic} should satisfy the condition $\vartriangle = b_p^2 - 4 a_p c_p \geq 0$: $\vartriangle=0$ is the critical condition for $p_{10}$ to have real solutions.
Noticing $\left(c_x - \lambda^2 \right) \left(c_x + \omega_p^2 \right)=-1$, the critical condition gives an ellipse of the form


\begin{equation}
	\frac{\left(\bar q_{10} \cos \vartheta + \bar q_{20} \sin \vartheta \right)^2}{a_e^2} +\frac{\left(- \bar q_{10} \sin \vartheta + \bar q_{20} \cos \vartheta \right)^2}{b_e^2} = 1,
	\label{R:standard-ellipse}
\end{equation}
where
\begin{equation*}
	\begin{split}
		& a_e= \sqrt{\frac{2 h \left(\lambda^2 + \omega_p^2\right)^2 \left(c_x + \omega_p^2 \right)^2}{\omega_p s_2^2 \left[ \left(c_x + \omega_p^2 \right)^2 + 1\right]}}, \hspace{0.7in} b_e= \sqrt{\frac{h \left(k_p - 1\right)^2 \left(\lambda^2 + \omega_p^2 \right)^2}{\lambda k_p s_1^2 \left[ \left(c_x + \omega_p^2 \right)^2 + 1\right]}},\\
		&  \cos \vartheta = \frac{1}{\sqrt{\left(c_x + \omega_p^2 \right)^2 + 1 }}, \hspace{1.3in}   \sin \vartheta= \frac{\left(c_x + \omega_p^2 \right)}{\sqrt{\left(c_x + \omega_p^2 \right)^2 + 1 } }\\
	\end{split}
\end{equation*}
The ellipse is counterclockwise tilted by $\vartheta$ from a standard ellipse $\bar q_{10}^2/a_e^2 + \bar q_{20}^2 /b_e^2=1$. The ellipse governed by \eqref{R:standard-ellipse} is the critical condition that $\bar p_{10}$ exists, so it is the boundary for asymptotic orbits. In other words, inside the ellipse, transit orbits exist, while outside the ellipse, transit orbits do not exist. As a result, we refer to the ellipse as the {\bf ellipse of transition} (see Figure \ref{lin_damp_position_paper}(b)). 
Note that on the boundary of the ellipse, there is only one asymptotic orbit (i.e., the wedge has collapsed into a single direction). 

The solutions to  \eqref{quadratic} are given by
\begin{equation}
	\bar p_{10}=\frac{-b_p \pm \sqrt{b_p^2 - 4 a_p c_p}}{2 a_p}
\end{equation}
and the expression for $\bar p_{20}$ is 
\begin{eqnarray}
	\bar p_{20}=\bar p_{10} \left(c_x + \omega_p^2 \right) + \frac{\lambda \left(1+ k_p \right) \left[\bar q_{20} -\bar q_{10} \left(c_x + \omega_p^2 \right) \right]}{k_p -1}
\end{eqnarray}
Up to now, the initial conditions $(\bar q_{10}, \bar q_{20}, \bar p_{10}, \bar p_{20})$ for the asymptotic orbits at a specific position have been obtained. The interior angle determined by these two initial velocities defines the wedge of velocites: $\theta =\arctan \left(\bar p_{20}/\bar p_{10} \right)$. 
The boundary of this wedge correspond to the asymptotic orbits. In fact, the wedge for the conservative system can be obtained by this method by taking $c_1$ as zero.

Figure \ref{lin_damp_position_paper} illustrates the projection on the configuration space in the equilibrium region. In the dissipative system, one important finding is the existence of the ellipse of transition given by \eqref{R:standard-ellipse}. The length of the major and minor axes of the ellipse are $a_e$ and $b_e$, respectively. For small damping, the major axis is much larger than the minor axis so that it reaches far beyond the neck region. Thus, here we  give the local flow near the neck region as shown in Figure \ref{lin_damp_position_paper}(a). We show a zoomed-out view revealing the entire ellipse in Figure \ref{lin_damp_position_paper}(b). The asymptotic orbits in the dissipative system are bounded by the ellipse (which is different from  the asymptotic orbits in the conservative system, which are bounded by the strip). Moreover, in the conservative system, all asymptotic orbits can reach the boundary of the strip with the period of $ 2 \pi / \omega_p$, while the asymptotic orbits in the dissipative system can never reach the boundary of the ellipse after they start due to damping. Notice that $a_e$ goes to zero when $c_1$ is large enough.  

Outside the ellipse, $\vartriangle= b_p^2 - 4 a_p c_p<0$, only non-transit orbits project onto this region. Thus we can conclude that the signs of $k_1 - k_3$ and $k_4 - k_2$ are independent of the direction of the velocity and can be determined in each of the components of the equilibrium region $R$ complementary to the ellipse. For example, in the left-most component, $k_1 - k_3$ is negative and $k_4 - k_2$ is positive, while in the right-most components, $k_1 - k_3$ is positive and $k_4 - k_2$ is negative.

Inside the ellipse the situation is more complex due to the existence of the wedge of velocity. For simplicity we still just show the wedges on the two vertical bounding line segments in Figure \ref{lin_damp_position_paper}. For example, consider the intersection of the strip with the left-most vertical line. At each position on the subsegment, one wedge of velocity exists in which $k_1 - k_2 $ is positive. The orbits with velocity interior to the wedge are transit orbits, and $k_4 - k_2$ is always positive. Orbits with velocity on the boundary of the wedge are asymptotic ($(k_1 - k_3) (k_4 - k_2)=0$), while orbits with velocity outside of the wedge are non-transit ($(k_1 - k_3) (k_4 - k_2)<0$). Notice that in Figure \ref{lin_damp_position_paper}, the grey shaded wedges are the wedges for the dissipative system, while the blacked shaded wedges partially covered by the grey ones are for conservative system (hardly visible for the parameters shown in the figure). The shrinking of the wedges from the conservative system to the dissipative system is caused by damping. This confirms the expectation that an increase in damping decreases the proportion of the transit orbits.

\section{Transition Tubes}

In this section, we go step by step through the numerical construction of the boundary between transit and non-transit orbits in the nonlinear system \eqref{eq:eomHam}.
We combine the geometric insight of the previous sections with numerical methods to demonstrate the existence of `transition tubes' for both the conservative and damped systems. Particular attention is paid to the modification of phase space transport as damping is increased, as this has not been considered previously.

\paragraph{Tube dynamics} 
The dynamic snap-through of the shallow arch can be understood as trajectories escaping from a potential well with energy above a critical level: the energy of the saddle point S$_1$. However, even if the energy of the system is higher than critical, the snap-through may not occur. The dynamical boundary between snap-through and non-snap-through behavior can be systematically understood by {\bf tube dynamics}. 
Tube dynamics  \cite{Conley1968, LlMaSi1985, OzDeMeMa1990, DeMeTo1991, DeLeon1992, Topper1997, KoLoMaRo2000, GaKoMaRo2005, GaKoMaRoYa2006, MaRo2006, KoLoMaRo2011} 
supplies a large-scale 
picture of transport; transport between the largest features of the phase space---the potential wells.  
In the conservative system, the stable and unstable manifolds with a $S^1 \times \mathbb{R}$ geometry act as {\bf tubes} emanating from the periodic orbits. 
While found above for the linearized system near S$_1$, these structures persist in the full nonlinear system
The manifold tubes (usually called {\bf transition tubes} in tube dynamics), formed by pieces of asymptotic orbits, separate two distinct types of orbits: transit orbits and non-transit orbits, corresponding to snap-through and non-snap-through in the present problem. The transit orbits, passing from one region to another through the bottleneck, are those inside the transition tubes. The non-transit orbits, bouncing back to their region of origin, are those outside the transition tubes. Thus, the transition tubes can mediate the global transport of states between snap-through and non-snap-through. 
In the dissipative system, similar transition tubes also exist.
Even in systems where stochastic effects are present, the influence of these structures remains \cite{NaRo2017}.

\subsection{Algorithm for computing transition tubes}

For  the conservative system, Ref.\ \cite{KoLoMaRo2011} gives a general numerical method to obtain the transition tubes. The key steps are (1) to find the periodic orbits restricted to a specified energy using differential correction and numerical continuation based on the initial conditions obtained from the linearized system at first, then (2) to compute the manifold tubes of the periodic orbits in the nonlinear system (i.e., `globalizing' the manifolds), and finally (3) to obtain the intersection of the Poincar\'{e} surface of section and global manifolds. See details in Ref.\ \cite{KoLoMaRo2011}. The method is effective in the conservative system, but not applicable to the dissipative system, since due to loss of conservation of energy, no periodic orbit exists. Thus, we provide another method  as follows.

\begin{figure}[!h]
\begin{center}
\includegraphics[width=0.7\textwidth]{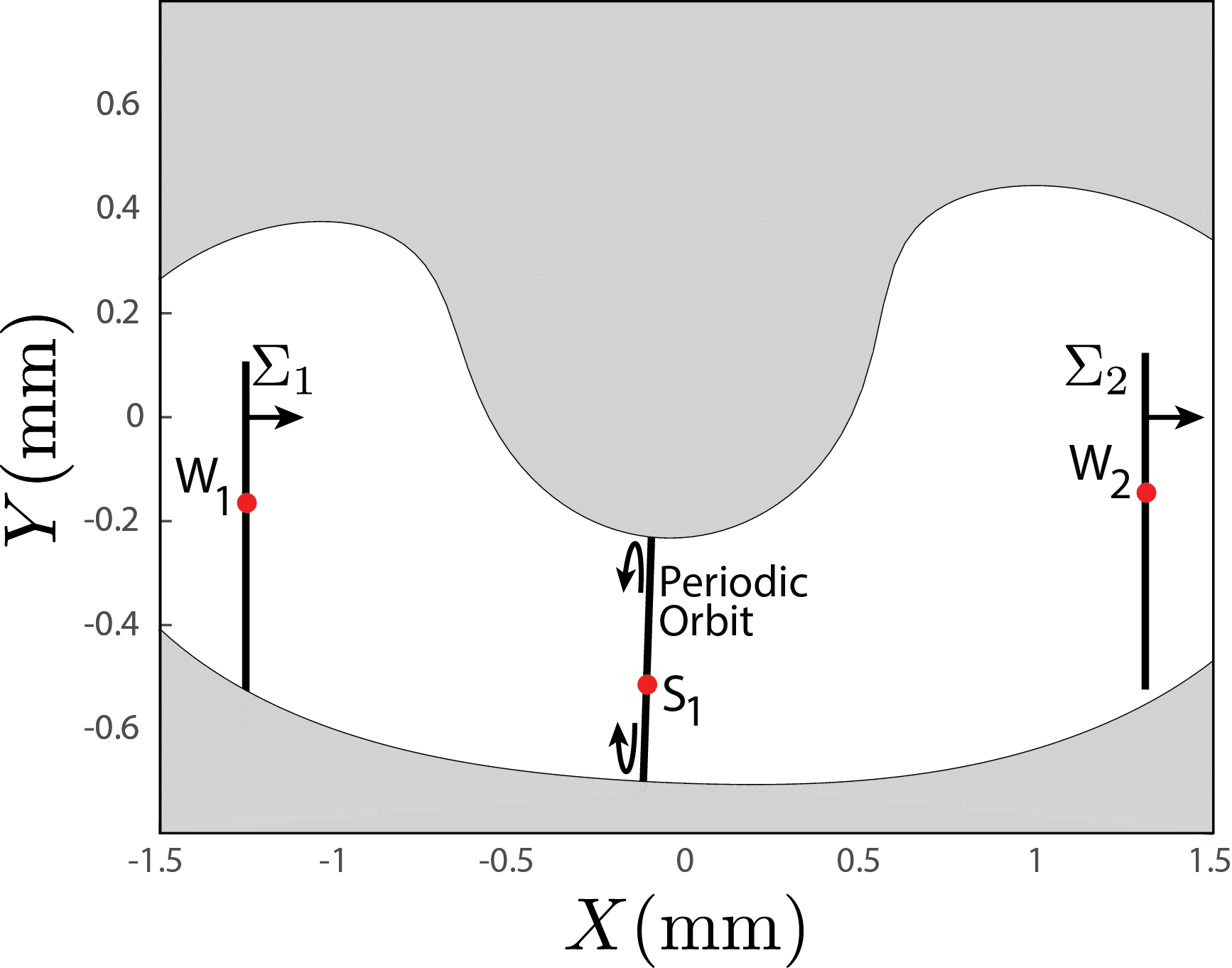} 
\caption{\label{non_sections_paper}{\footnotesize 
For a representative energy above the saddle point S$_1$, we show the unstable periodic orbit in the neck region around S$_1$. It projects to a single line going between the upper and lower energy boundary curves, and arrows are shown for convenience.
 We show the Poincar\'e sections $\Sigma_1$ and $\Sigma_2$ which are defined by $X$ values equal to that of the two stable equilibria in the center of the left and right side wells, W$_1$ and W$_2$, respectively. The arrows on the vertical lines indicate that these Poincar\'e sections are also defined by positive $X$ momentum.
}}
\end{center}
\end{figure}

{\bf Step 1: Select an appropriate energy.} 
We first need to set the energy to an appropriate value such that the snap-though behavior exists.  Once the energy is given,  it remains constant in the conservative system. In our example, the critical energy for snap-through is the energy of S$_1$. Thus, we can choose an energy which is between that of S$_1$ and S$_2$. In this case, all transit orbits can just escape from W$_1$ to W$_2$ through S$_1$. Notice that the potential energy determines the width of the bottleneck and the size of the transition tubes which hence determines the relative fraction of transit orbits in the phase space. A representative energy case is shown in Figure \ref{non_sections_paper}, which also establishes our location for Poincar\'e sections $\Sigma_1$ and $\Sigma_2$ which are at $X=$constant lines passing through W$_1$ and W$_2$ respectively, and with $p_X>0$.

{\bf Step 2: Compute the approximate transition tube and its intersection on a Poincar\'{e} section.} 
We have analyzed the flow of linearized system in both phase space and position space which classifies orbits into four classes. It shows that in the conservative system the stable manifolds correspond to the boundary between transit orbits and non-transit orbits. Thus, we can choose this manifold as the starting point. We start by considering the approximation of transition tubes for the conservative system.

{\it Determine the initial condition.}
The stable manifold divides the transit orbits and non-transit orbits for all trajectories headed toward a bottleneck. Thus, we can use the stable manifold to obtain the initial condition.  Considering the general solutions  \eqref{whole-gener-sol} of the linearized equations  \eqref{conservative eqns}, we can let $p_1^0=c$, $q_1^0=0$, $q_2^0=A_q$ and $p_2^0=A_p$. Notice that
\begin{equation}
A_q^2+A_p^2=2 h/\omega_p
\label{circle}
\end{equation} 
which forms a circle in the center projection, so in the next computational procedure we should pick up $N$ points on the circle with a constant arc length interval. Each $A_q$ and $A_p$ determined by these sampling points along with $p_1^0=c$ and $q_1^0=0$ can be used as initial conditions. When first transformed back to the position space and then transformed to dimensional quantities, this yields an initial condition
\begin{equation}
\begin{pmatrix}
X_0\\ Y_0\\p_{X0} \\ p_{Y0}
\end{pmatrix} 
=
\begin{pmatrix}
x_e\\ y_e\\0\\ 0
\end{pmatrix} + \begin{pmatrix}
L_x\\ L_y\\ \omega_0 L_x M_1\\ \omega_0 L_y M_2
\end{pmatrix}^T
C
\begin{pmatrix}
c\\0\\A_q\\A_p
\end{pmatrix}
\label{Initial}
\end{equation}

{\it Integrate backward and obtaining Poincar\'{e} section.}
Using the $N$ initial conditions \eqref{Initial} yielded by varying $A_q$ and $A_p$ governed by \eqref{circle} and integrating the nonlinear equations of motions in \eqref{eq:eomHam} in the backward direction, we  obtain a tube, formed by the $N$ trajectories, which is a linear approximation for the transition tube. Choosing the Poincar\'{e} surface-of-section $\Sigma_1$ is shown in Figure \ref{non_sections_paper}, corresponding to $X=X_{{\rm W}_1}$ and $p_X>0$.

{\bf Step 3: Compute the real transition tube by the bisection method.} 
We have obtained a Poincar\'{e} section which is the intersection of the approximate transition tube and the surface $\Sigma_1$. First pick a point (noted as $p_i$) which is almost the center of the closed curve. The line from $p_i$ to each of the $N$ points on the Poincar\'{e} map will form a ray. The point $p_i$ inside the curve in general is a transit orbit. Then choose another point on each radius which is a non-transit orbit, noted as $p_o$. With the approach described above, we can use the bisection method to obtain the boundary of the transition tube on a specific radius (cf.\ \cite{AnEaLo2017}). Picking the midpoint (marked by $p_m$) as the initial condition and carrying out forward integration for the nonlinear equation of motion in \eqref{eq:eomHam}, we can estimate if this trajectory can transit or not. If it is a transit orbit, note it as $p_i$, or note it as $p_o$. 
Continuing this procedure again until the distance between $p_i$ and $p_o$ 
reaches a specified tolerance, the boundary of the tube on this ray is estimated. Thus, the real transition tube for the conservative system can be obtained if the same procedure is carried out for all angles. 
A related method is described in \cite{OnYoRo2017}, adapting an approach of \cite{GaMaDuCa2009}.

For the dissipative system, the size of the transition tubes for a given  energy on $\Sigma_1$ will shrink. Using the bisection method and following the same procedure  as for conservative system, the transition tube for the dissipative system will be obtained.

\subsection{Numerical results and discussion}

To visualize the tube dynamics for the arch, several examples will be given. According to the steps mentioned above, we can obtain the transition tubes for both the conservative system and dissipative systems. 
For all results, the geometries of the arch are selected as $b=12.7$ mm $d=0.787$ mm, $L=228.6$ mm. 
The Young's modulus and the mass density are $E=153.4$ GPa and $\rho=7567 \ \mathrm{kg \ m^{-3}}$. The selected thermal load corresponds to $184.1$ N, while the initial imperfections are $\gamma_1 = 0.082$ mm and $\gamma_2 = -0.077$ mm. 
These values match the parameters given in the experimental study \cite{WiVi2016}.
For all the numerical results given in this section, the initial energy of the system is set at
3.68$\times 10^{-4}$ J - above the energy of saddle point S$_1$, so that the equilibrium point $W_1$ is inside the configuration space projection. This choice of initial energy will make it possible to compare the numerical results with the experimental results which are planned for future work. 

\begin{figure}[!th]
\begin{center}
\includegraphics[width=\textwidth]{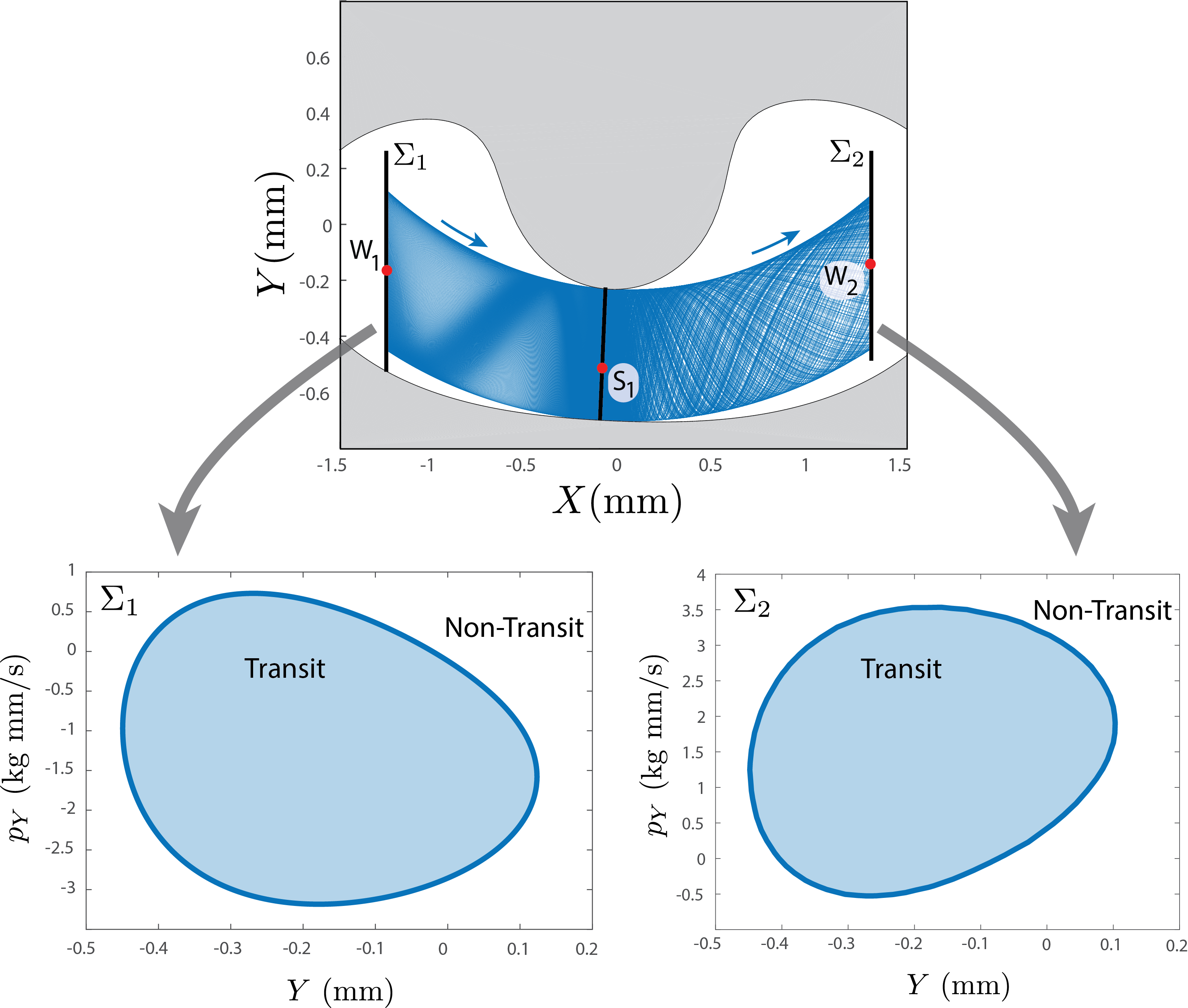} 
\caption{\label{non_conservative_tube_all_paper}{\footnotesize 
A transition tube from the left well to the right well, obtained using the method described in the text. The upper figure shows the configuration space projection. 
The lower left shows the tube boundary (closed curve) on Poincar\'e section $\Sigma_1$, which separates transit and non-transit trajectories.
The lower right shows the corresponding tube boundary (closed curve) on Poincar\'e section $\Sigma_2$. 
}}
\end{center}
\end{figure}

\paragraph{Transition tubes for conservative system}   
For conservative system, the Hamiltonian is a constant of motion. 
In Figure \ref{non_conservative_tube_all_paper}, we show the configuration space projection of the transition tube and the Poincar\'{e} sections on $\Sigma_1$ and $\Sigma_2$ which are closed curves. In Figure \ref{non_conservative_tube_all_paper} are shown all the trajectories which form the transition tube boundary starting from $\Sigma_1$ and ending up at $\Sigma_2$, flowing from left to right through the neck region. 

Due to the the conservation of energy, the size of the transition tube is constant during evolution, which corresponds to the cross-sectional area of the transition tube. It should be noted that the areas of the tube Poincar\'{e} sections on $\Sigma_1$ and $\Sigma_2$  in Figure \ref{non_conservative_tube_all_paper}
are equal, due to the integral invariants of Poincar\'e for a system obeying Hamilton's canonical equations (with no damping).
Moreover, note that the size of the transition tube, the boundary of the transit orbits, is determined by the energy.
For a lower energy, the size of the transition tube is smaller or vice versa. 
In other words, the area of the Poincar\'{e} sections on $\Sigma_1$ and $\Sigma_2$ is determined by the energy. In fact, the cross-sectional area of the transition tube is proportional to the energy above the saddle point S$_1$ \cite{MacKay1990}. As mentioned before, the transition tube separates the transit orbits and non-transit orbits, which correspond to snap-through and non-snap-through. The orbit inside the transition tube can transit, while the orbit outside the transition tube cannot transit.

\paragraph{Transition tubes for dissipative system}  
Unlike the conservation of energy in conservative system, the energy in the dissipative system is decreasing with time.
\begin{figure}[!th]
	\begin{center}
		\includegraphics[width=\textwidth]{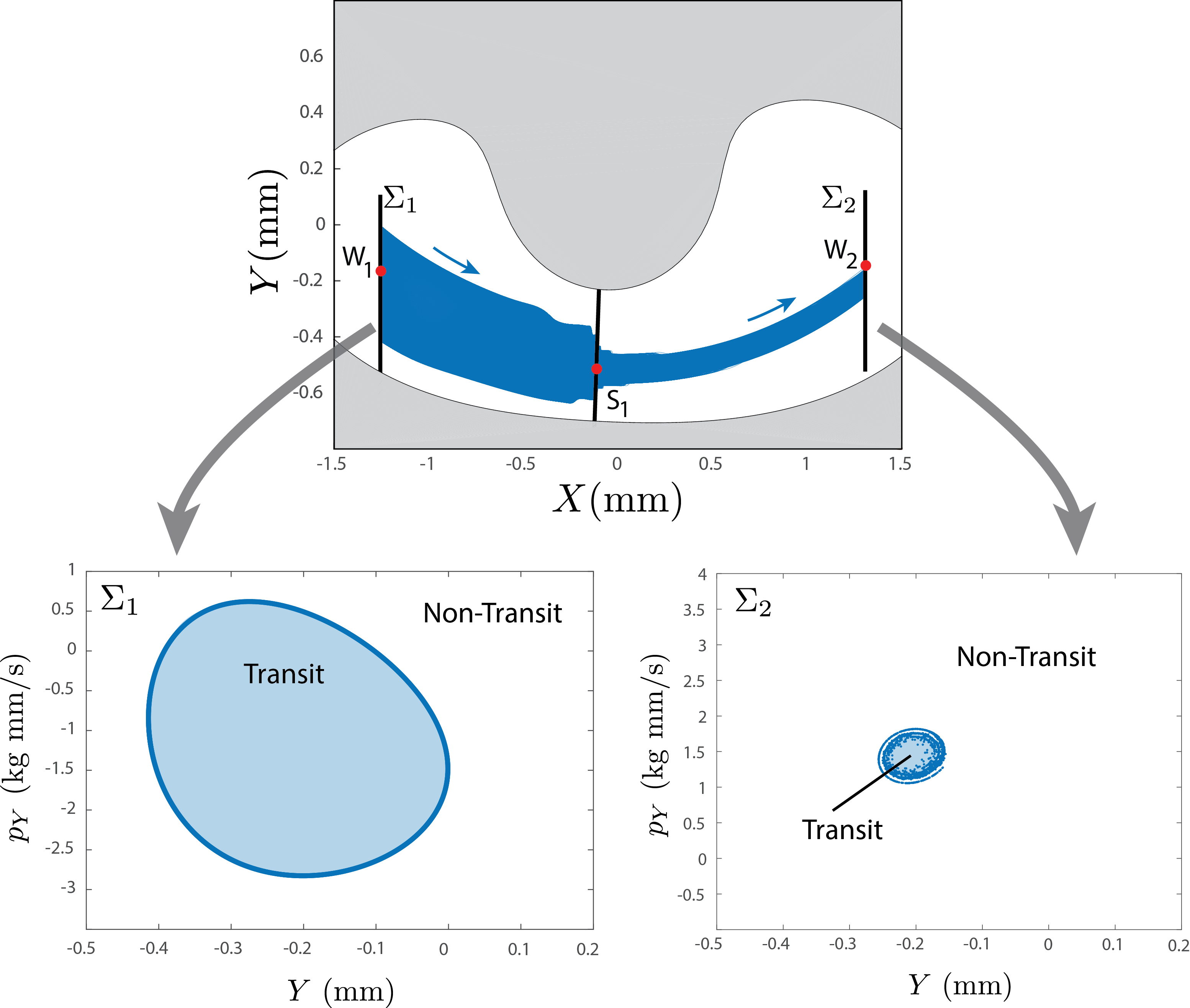} 
		\caption{\label{non_damp_tube_all_paper}{\footnotesize 
				A transition tube from the left well to the right well, obtained using the method described in the text, for the case of damping. The upper figure shows the configuration space projection. 
				The lower left shows the tube boundary (closed curve) on Poincar\'e section $\Sigma_1$ which separates transit and non-transit trajectories for initial conditions all with a given fixed initial energy.
				The lower right shows the corresponding image under the flow on Poincar\'e section $\Sigma_2$. 
				Due to the damping, and a range of times spent in the neck region, spiraling is visible in this 2D projection since trajectories which spend longer in the neck will be at lower total energies. Compare with Figure \ref{non_conservative_tube_all_paper}.
			}}
		\end{center}
\end{figure}
Figure \ref{non_damp_tube_all_paper} 
shows the configuration space projection of the transition tube and the Poincar\'{e} sections on $\Sigma_1$ and $\Sigma_2$.  
In Figure \ref{non_damp_tube_all_paper} the transition tube starts from $\Sigma_1$ and ends up with $\Sigma_2$ flowing from left to right through the neck region, as shown previously for the conservative system. 
From the figure, we can observe the distinct reduction in the size of the transition tube, especially near the neck region.
To show this, the scale of the Poincar\'e section projections is the same as in Figure \ref{non_conservative_tube_all_paper}. 
During the evolution, the energy of the system is decreasing due to damping. The trajectories spend a great amount of time  crossing the neck region, resulting in  the total energy decreasing dramatically (and influencing the size of the transition tubes to the right of the neck region). Thus, the transition tube is spiraling in the neck region so that Poincar\'{e} $\Sigma_2$ is not a closed curve, nor are the trajectories at a constant energy.
The  $\Sigma_2$ plot is merely a projection onto the $(Y,p_Y)$-plane to give an idea of the actual co-dimension 1 tube boundary in the 4-dimensional phase space.
Note the clear differences between Figure \ref{non_conservative_tube_all_paper} and Figure \ref{non_damp_tube_all_paper}. 
The dramatic shrinking of tubes near the neck region is due almost entirely to the linearized dynamics near the saddle point.  To confirm this, we present the linear transition tube obtained by the analytical solutions  \eqref{Hd gener solut} for the linearized dissipative system in Figure \ref{lin_damp_tube_paper}. 

\begin{figure}[!h]
	\begin{center}
		\includegraphics[width=0.7\textwidth]{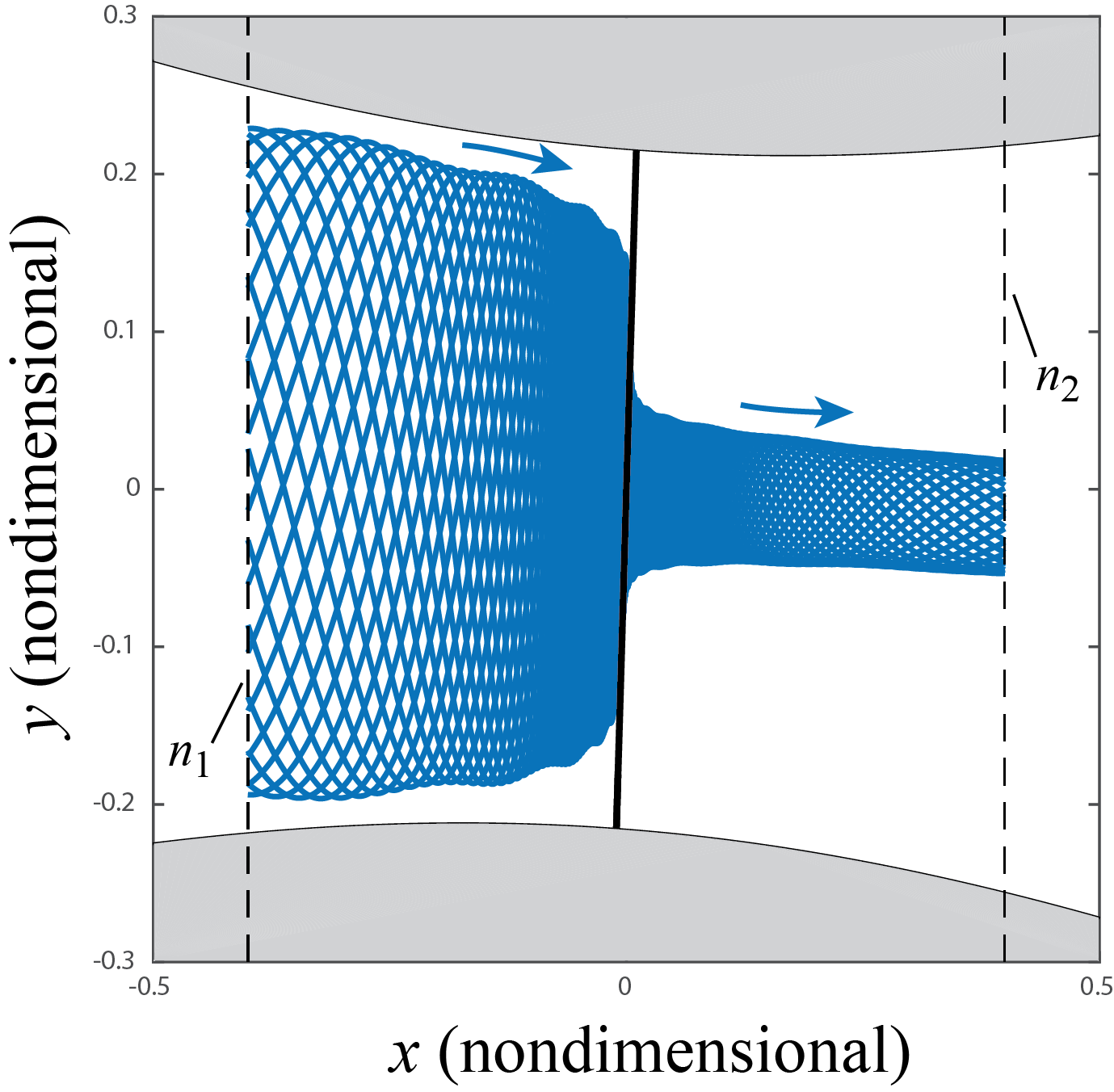} 
		\caption{\label{lin_damp_tube_paper}{\footnotesize 
				A transition tube from the left side boundary ($n_1$) to the right side boundary ($n_2$) of the equilibrium region around saddle point S$_1$, obtained for the linear damped system. 
				Notice that the shrinking of the tube is observed as in the nonlinear system, Figure \ref{non_damp_tube_all_paper}, here seen in terms of the width of the projected strip onto configuration space.
			}}
		\end{center}
	\end{figure}

\begin{figure}[!h]
	\begin{center}
		\includegraphics[width=\textwidth]{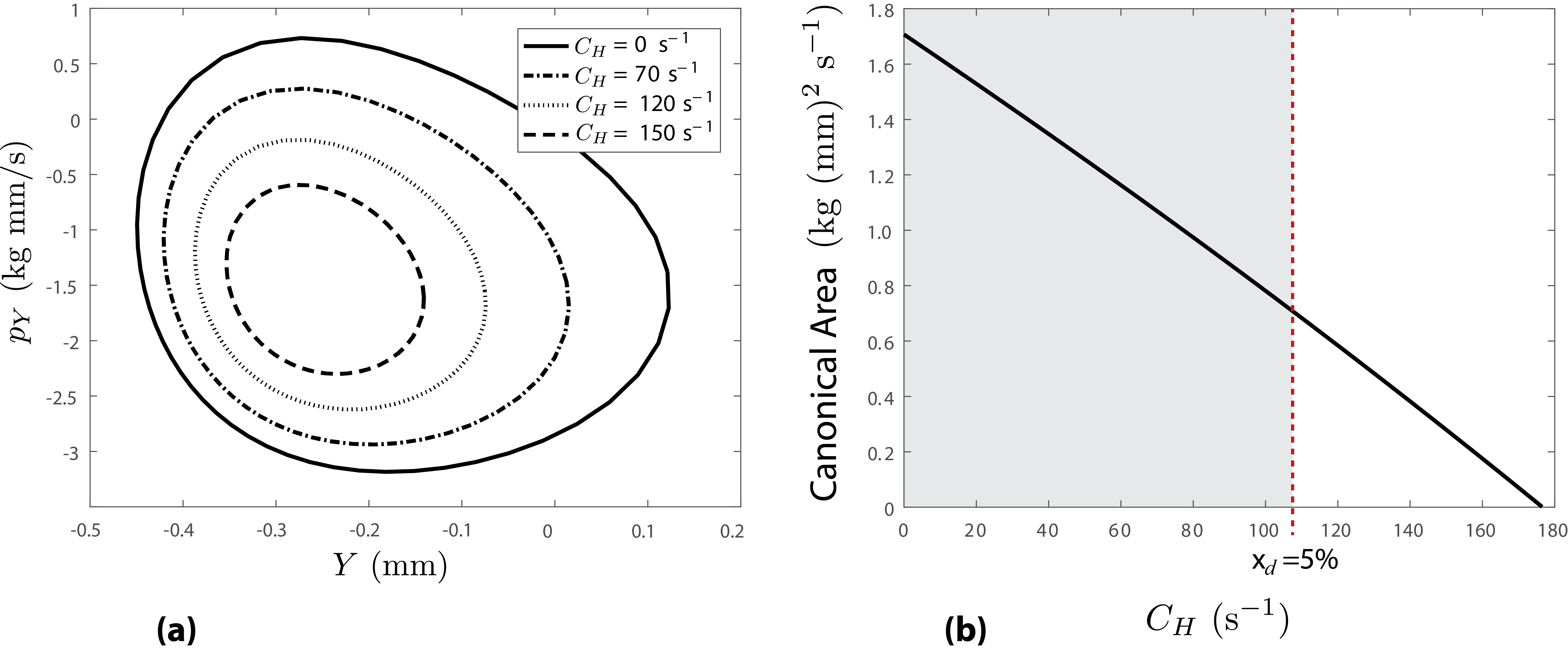} 
		\caption{\label{multi_damping_Poincare_both_paper}{\footnotesize 
				The effect of the damping coefficient $C_H$ on the area of the transition tube on Poincar\'e section $\Sigma_1$ is shown. For a fixed initial energy 
				above the saddle, the projection on the canonical plane $(Y,p_Y)$ is shown in (a) and the area is plotted in (b). 
				In (b), the shaded region indicates the experimentally observed range of damping coefficients, which correspond to non-dimensional damping factor $\xi_d$ less than 5\%.
			}}
		\end{center}
\end{figure}

\paragraph{Effect of damping on the transition tubes} 
In order to further quantify how  damping affects the size of  transition tubes, we present the tube Poincar\'{e} section on  $\Sigma_1$ with different damping in Figure \ref{multi_damping_Poincare_both_paper}. 
In Figure \ref{multi_damping_Poincare_both_paper}(a), we can see the canonical area ($\int_\mathcal{A} p_Y dY$) decreases with increasing damping. Thus, the proportion of transition trajectories will be fewer if the damping increases. Note that when the damping changes, different transition tubes almost share the same center which corresponds to the fastest trajectories. 
Figure \ref{multi_damping_Poincare_both_paper}(b) shows the relation between the damping and the projected canonical area ($\int_\mathcal{A} p_Y dY$), which is related to the relative number of transit compared to non-transit orbits. 
It shows that an increase in damping decreases the projected area.
When the damping is small, the relation between the damping and the  area is linear, while when the damping is large, the relation becomes slightly nonlinear.
Note that generally in mechanical/structural experiments the non-dimensional damping factor $\xi_d$ is less than $5\%$ which corresponds to a damping coefficient $C_H$ less than $107.3 \  \mathrm{s^{-1}}$ (see the shaded region in Figure \ref{multi_damping_Poincare_both_paper}(b)).
Furthermore, note that for the initial energy depicted in Figure \ref{multi_damping_Poincare_both_paper}, there are are no transit orbits starting on 
$\Sigma_1$ for  $C_H$ greater than about $185 \  \mathrm{s^{-1}}$.

\begin{figure}[!h]
	\begin{center}
		\includegraphics[width=\textwidth]{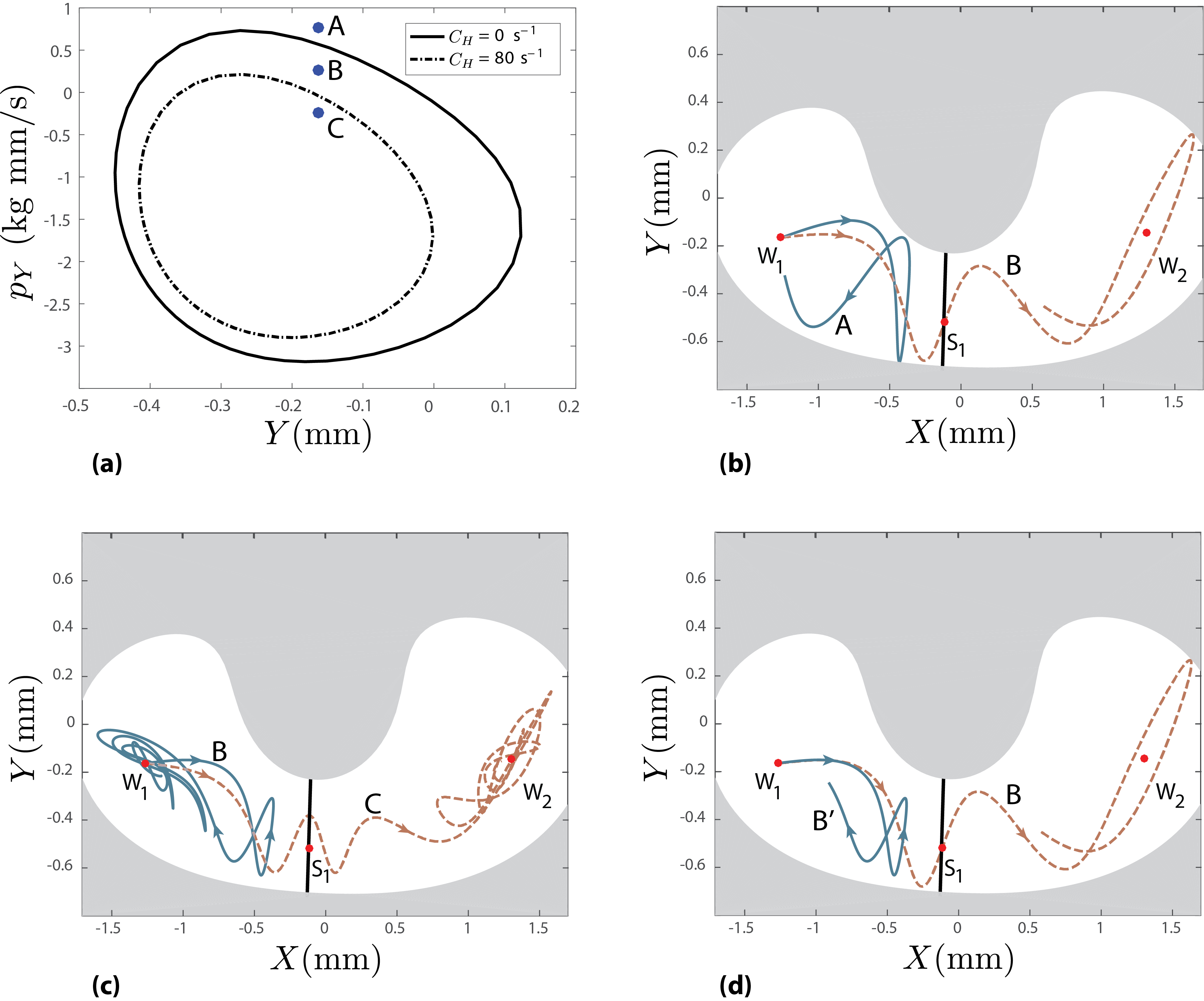} 
		\caption{\label{time-history-Poincare-all-mod-paper}{\footnotesize 
				Several example trajectories are shown, starting from the stable well point W$_1$. The initial conditions from Poincar\'e section $\Sigma_1$ are shown in (a) for a fixed initial energy, along with the transition tube boundaries for the conservative case and a damped case. In (b), we show the trajectories for points A and B, for the conservative case where A is just outside the tube boundary and B is just inside.
				In (c), we show the 
				trajectories for points B and C, for the damped case where B is just outside the tube boundary and C is just inside.
				In (d), we illustrate the effect of damping by starting the same initial condition, B, but showing the trajectory in the conservative case as trajectory B and the damped case as trajectory B$^{\prime}$. 
			}}
		\end{center}
\end{figure}

\paragraph{Demonstration of trajectories inside and outside the transition tube} 
To illustrate the effectiveness of the transition tubes, we choose three points on $\Sigma_1$ (see A, B and C in Figure \ref{time-history-Poincare-all-mod-paper}(a)) as the initial conditions and integrate forward to see their evolution. 
Note that all the trajectories corresponding to these three points have the same initial energy and start from a configuration identical to  the equilibrium point $W_1$, but with different initial velocity directions.
Figure \ref{time-history-Poincare-all-mod-paper}(b) shows the trajectories A and B in the conservative system where A is outside the tube boundary and B is inside the tube boundary. 
In the figure, trajectory B transits through the neck region and trajectory A bounces back. 
Figure \ref{time-history-Poincare-all-mod-paper}(c) shows trajectories B and C in the dissipative system. Like the situation in the conservative system, trajectory C which is inside the tube can transit, while trajectory B which is outside the tube cannot. 
Figure \ref{time-history-Poincare-all-mod-paper}(d) shows the effect of damping on the transit condition for the trajectories B and B$^{\prime}$ with the same initial condition. Trajectory B is simulated using the conservative system and trajectory B$^{\prime}$ is simulated using the dissipative system. It shows that the damping changes the transit condition that a transit orbit B in the conservative system becomes non-transit orbit B$^{\prime}$ in the dissipative system, both starting from the same initial condition. 
From Figure \ref{time-history-Poincare-all-mod-paper}, we can conclude the transition tube can effectively estimate the snap-through transitions both in conservative systems and dissipative systems.

Finally, we point out that the transition tubes are the boundary for transit orbits that transition {\it the first time}. For example,  trajectory A in Figure \ref{time-history-Poincare-all-mod-paper}(b) stays outside of the transition tube so that it returns near the neck region at first, but, unless it happens to be on a KAM torus or a stable manifold of such a torus, it will ultimately transit  as long as it does not form a periodic orbit near the potential well W$_1$, since the energy is above the critical energy for transition and is conservative.

\section{Conclusions}

Tube dynamics  is a conceptual dynamical systems framework initially used to study the isomerization reactions in chemistry \cite{OzDeMeMa1990, DeMeTo1991, DeLeon1992, Topper1997, JaFaUz1999} as well as other fields, like resonance transitions in
celestial mechanics 
\cite{LlMaSi1985, KoLoMaRo2000, JaRoLoMaFaUz2002, GaKoMaRoYa2006, MaRo2006}
and capsize in ship dynamics \cite{NaRo2017}. 
Here we extend the application of tube dynamics to structural mechanics:\ the snap-through of a shallow arch, or buckled-beam. In general, slender elastic structures are capable of exhibiting a variety of (co-existing) equilibrium shapes, and thus, given a disturbance, tube dynamics sheds light on how such a system might be caused to transition between available, stable equilibrium configurations. Moreover, it is the first time, to the best of our knowledge, that tube dynamics has been worked out for a dissipative system, which increases the generality of the approach.


The snap-through transition of an arch was studied via a two-mode truncation of the governing partial differential equations based on Euler-Bernoulli beam theory. Via analysis of the linearized Hamiltonian equations around the saddle, the analytical solutions for both the conservative and dissipative systems were determined and the corresponding flows in the equilibrium region of eigenspace and configuration space were discussed. The results show that all transit orbits, corresponding to snap-through, must evolve from a wedge of velocities which are restricted to a strip in configuration space in the conservative system, and by an ellipse in the corresponding dissipative system when damping is included. Using the results from the linearization as an approximation, the transition tubes based on the full nonlinear equations for both the conservative  and dissipative system were obtained by the bisection method. The orbits inside the transition tubes can transit, while the orbits outside the tubes cannot. Results also show that the damping makes the size of the transition tubes smaller, which corresponds to the degree, or amount, of orbits that transit. When the damping is small, it has a nearly linear effect on the size of the transition tubes. 

Further study of the dynamic behaviors of the arch can lead to more immediate application structural mechanics. For example, many structural systems possess multiple equilibria, and the manner in which the governing potential energy changes with a control parameter is, of course, the essence of bifurcation theory. However, under nominally fixed conditions, the present paper directly assesses the energy required to (dynamically) perturb a structural system beyond the confines of its immediate potential energy well. In future work, a three-mode truncation may be introduced to study such systems. High order approximations will present higher index saddles  which will modify the tube dynamics framework presented here (cf.\ \cite{collins2011index} \cite{HallerUzer2011}  \cite{Nagahata2013}). Furthermore, experiments will be carried out to show the effectiveness of the present approach to prescribe initial conditions which lead to dynamic buckling.

\section{Acknowledgements}
This work was supported in part by the National Science Foundation under awards 1150456 (to SDR) and 1537349 (to SDR and LNV). 
One of the authors (SDR) acknowledges enjoyable interactions during the past  decade with Professor Romesh Batra, who is being honored by this issue.



\begin{thebibliography}{10}
\providecommand{\url}[1]{{\tt #1}}
\providecommand{\urlprefix}{URL }

\bibitem{WiVi2016}
Wiebe, R. and Virgin, L.~N. [2016] On the experimental identification of
  unstable static equilibria.
\newblock {\em Proceedings of the Royal Society of London A: Mathematical,
  Physical and Engineering Sciences\/} {\bf 472}(2190):20160172.

\bibitem{collins2012isomerization}
Collins, P., Ezra, G.~S. and Wiggins, S. [2012] Isomerization dynamics of a
  buckled nanobeam.
\newblock {\em Physical Review E\/} {\bf 86}(5):056218.

\bibitem{virgin2017geometric}
Virgin, L., Guan, Y. and Plaut, R. [2017] On the geometric conditions for
  multiple stable equilibria in clamped arches.
\newblock {\em International Journal of Non-Linear Mechanics\/} {\bf 92}:8--14.

\bibitem{das2009symmetry}
Das, K. and Batra, R. [2009] Symmetry breaking, snap-through and pull-in
  instabilities under dynamic loading of microelectromechanical shallow arches.
\newblock {\em Smart Materials and Structures\/} {\bf 18}(11):115008.

\bibitem{das2009pull}
Das, K. and Batra, R. [2009] Pull-in and snap-through instabilities in
  transient deformations of microelectromechanical systems.
\newblock {\em Journal of Micromechanics and Microengineering\/} {\bf
  19}(3):035008.

\bibitem{mann2009energy}
Mann, B. [2009] Energy criterion for potential well escapes in a bistable
  magnetic pendulum.
\newblock {\em Journal of Sound and Vibration\/} {\bf 323}(3):864--876.

\bibitem{Thompson1984}
Thompson, J. M.~T. and Hunt, G.~W. [1984] {\em Elastic Instability
  Phenomena\/}.
\newblock Wiley.

\bibitem{NaRo2017}
Naik, S. and Ross, S.~D. [2017] Geometry of escaping dynamics in nonlinear ship
  motion.
\newblock {\em Communications in Nonlinear Science and Numerical Simulation\/}
  {\bf 47}:48 -- 70.

\bibitem{KoLoMaRo2000}
Koon, W.~S., Lo, M.~W., Marsden, J.~E. and Ross, S.~D. [2000] Heteroclinic
  connections between periodic orbits and resonance transitions in celestial
  mechanics.
\newblock {\em Chaos\/} {\bf 10}:427--469.

\bibitem{Conley1968}
Conley, C.~C. [1968] Low energy transit orbits in the restricted three-body
  problem.
\newblock {\em SIAM J. Appl. Math.\/} {\bf 16}:732--746.

\bibitem{LlMaSi1985}
Llibre, J., Martinez, R. and Sim\'o, C. [1985] Transversality of the invariant
  manifolds associated to the {L}yapunov family of periodic orbits near {L}2 in
  the restricted three-body problem.
\newblock {\em J. Diff. Eqns.\/} {\bf 58}:104--156.

\bibitem{OzDeMeMa1990}
{Ozorio de Almeida}, A.~M., {De Leon}, N., Mehta, M.~A. and Marston, C.~C.
  [1990] Geometry and dynamics of stable and unstable cylinders in
  {H}amiltonian systems.
\newblock {\em Physica D\/} {\bf 46}:265--285.

\bibitem{DeMeTo1991}
{De Leon}, N., Mehta, M.~A. and Topper, R.~Q. [1991] Cylindrical manifolds in
  phase space as mediators of chemical reaction dynamics and kinetics. {I}.
  {T}heory.
\newblock {\em J. Chem. Phys.\/} {\bf 94}:8310--8328.

\bibitem{DeLeon1992}
{De Leon}, N. [1992] Cylindrical manifolds and reactive island kinetic theory
  in the time domain.
\newblock {\em J. Chem. Phys.\/} {\bf 96}:285--297.

\bibitem{Topper1997}
Topper, R.~Q. [1997] Visualizing molecular phase space: nonstatistical effects
  in reaction dynamics.
\newblock In {\em Reviews in Computational Chemistry\/} (edited by K.~B.
  Lipkowitz and D.~B. Boyd), vol.~10, chap.~3, 101--176. VCH Publishers, New
  York.

\bibitem{GaKoMaRo2005}
Gabern, F., Koon, W.~S., Marsden, J.~E. and Ross, S.~D. [2005] Theory and
  Computation of Non-{RRKM} Lifetime Distributions and Rates in Chemical
  Systems with Three or More Degrees of Freedom.
\newblock {\em Physica D\/} {\bf 211}:391--406.

\bibitem{GaKoMaRoYa2006}
Gabern, F., Koon, W.~S., Marsden, J.~E., Ross, S.~D. and Yanao, T. [2006]
  Application of tube dynamics to non-statistical reaction processes.
\newblock {\em Few-Body Systems\/} {\bf 38}:167--172.

\bibitem{MaRo2006}
Marsden, J.~E. and Ross, S.~D. [2006] New methods in celestial mechanics and
  mission design.
\newblock {\em Bulletin of the American Mathematical Society\/} {\bf
  43}:43--73.

\bibitem{KoLoMaRo2011}
Koon, W.~S., Lo, M.~W., Marsden, J.~E. and Ross, S.~D. [2011] {\em Dynamical
  Systems, the Three-Body Problem and Space Mission Design\/}.
\newblock Marsden Books, ISBN 978-0-615-24095-4.

\bibitem{Murphy1996}
Murphy, K.~D., Virgin, L.~N. and Rizzi, S.~A. [1996] Experimental snap-through
  boundaries for acoustically excited, thermally buckled plates.
\newblock {\em Experimental Mechanics\/} {\bf 36}:312--317.

\bibitem{Wiebe2013}
Wiebe, R., Virgin, L.~N., Stanciulescu, I., Spottswood, S.~M. and Eason, T.~G.
  Characterizing Dynamic Transitions Associated with Snap-Through: A Discrete
  System.
\newblock {\em Journal of Computational and Nonlinear Dynamics\/} {\bf 8}.

\bibitem{zhong2016analysis}
Zhong, J., Fu, Y., Chen, Y. and Li, Y. [2016] Analysis of nonlinear dynamic
  responses for functionally graded beams resting on tensionless elastic
  foundation under thermal shock.
\newblock {\em Composite Structures\/} {\bf 142}:272--277.

\bibitem{Greenwood2003}
Greenwood, D.~T. [2003] {\em Advanced Dynamics\/}.
\newblock Cambridge University Press.

\bibitem{Wiggins1994}
Wiggins, S. [1994] {\em Normally Hyperbolic Invariant Manifolds in Dynamical
  Systems\/}.
\newblock Springer-Verlag, New York.

\bibitem{AnEaLo2017}
Anderson, R.~L., Easton, R.~W. and Lo, M.~W. [2017] Isolating blocks as
  computational tools in the circular restricted three-body problem.
\newblock {\em Physica D: Nonlinear Phenomena\/} {\bf 343}:38 -- 50.

\bibitem{OnYoRo2017}
Onozaki, K., Yoshimura, H. and Ross, S.~D. [2017] Tube dynamics and low energy
  {E}arth-{M}oon transfers in the 4-body system.
\newblock {\em Advances in Space Research\/} (~):to~appear.

\bibitem{GaMaDuCa2009}
Gawlik, E.~S., Marsden, J.~E., {Du {T}oit}, P.~C. and Campagnola, S. [2009]
  Lagrangian coherent structures in the planar elliptic restricted three-body
  problem.
\newblock {\em Celestial Mechanics and Dynamical Astronomy\/} {\bf
  103}:227--249.
\newblock {\tt doi:10.1007/s10569-008-9180-3}.

\bibitem{MacKay1990}
MacKay, R.~S. [1990] Flux over a saddle.
\newblock {\em Physics Letters A\/} {\bf 145}:425--427.

\bibitem{JaFaUz1999}
Jaff\'e, C., Farrelly, D. and Uzer, T. [1999] Transition state in atomic
  physics.
\newblock {\em Phys. Rev. A\/} {\bf 60}:3833--3850.

\bibitem{JaRoLoMaFaUz2002}
Jaff\'e, C., Ross, S.~D., Lo, M.~W., Marsden, J.~E., Farrelly, D. and Uzer, T.
  [2002] Theory of asteroid escape rates.
\newblock {\em Physical Review Letters\/} {\bf 89}:011101.

\bibitem{collins2011index}
Collins, P., Ezra, G.~S. and Wiggins, S. [2011] Index k saddles and dividing
  surfaces in phase space with applications to isomerization dynamics.
\newblock {\em The Journal of chemical physics\/} {\bf 134}(24):244105.

\bibitem{HallerUzer2011}
Haller, G., Uzer, T., Palacian, J., Yanguas, P. and Jaffe, C. [2011] Transition
  state geometry near higher-rank saddles in phase space.
\newblock {\em Nonlinearity\/} {\bf 24}(2):527.

\bibitem{Nagahata2013}
Nagahata, Y., Teramoto, H., Li, C.-B., Kawai, S. and Komatsuzaki, T. [2013]
  Reactivity boundaries for chemical reactions associated with higher-index and
  multiple saddles.
\newblock {\em Phys. Rev. E\/} {\bf 88}:042923.

\end{thebibliography}
\end{document}